\documentclass[twocolumngrid,prd,superscriptsize,reprint]{revtex4-1}
\usepackage[utf8]{inputenc} 
\usepackage{graphicx,graphics,xcolor,overpic,mathtools}
\usepackage{amsthm,amsmath,amssymb}
\usepackage[colorlinks]{hyperref}
\usepackage{textcomp}
\definecolor{coolblack}{rgb}{0.0, 0.18, 0.39}
\hypersetup{
    colorlinks = false,
   }
\PassOptionsToPackage{normalem}{ulem}

\usepackage{ulem} 
\usepackage{bm}
\usepackage{url}
\usepackage{physics}
\usepackage[makeroom]{cancel}
\usepackage{cleveref}
\usepackage{todonotes}
\usepackage{tensor}
\usepackage{mathrsfs}
\usepackage{slashed}
\usepackage[mode=buildnew]{standalone}

\newcommand{\comment}[1]{}

\usepackage{pgfplots}

\usepackage{diagbox}
\usepackage{stackengine,amssymb,graphicx}

\NewDocumentCommand{\evat}{sO{\bigg}mm}{%
  \IfBooleanTF{#1}
   {\mleft. #3 \mright|_{#4}}
   {#3#2|_{#4}}%
}
\usepackage{enumerate}
\definecolor{azure}{rgb}{0.0, 0.5, 1.0}

\begin{document}

\title[]{Rotating fermion-boson stars}

\author{Jorge Castelo Mourelle}
\author{Christoph Adam}
\author{Juan Calder\'on Bustillo }

\affiliation{%
Departamento de F\'isica de Part\'iculas, Universidad de Santiago de Compostela and Instituto
Galego de F\'isica de Altas Enerxias (IGFAE), E-15782 Santiago de Compostela, Spain
}%
\author{Nicolas Sanchis-Gual}

\affiliation{
Departamento de Astronom\'ia y Astrof\'isica, Universitat de Val\`encia,
Dr. Moliner 50, 46100 Burjassot (Val\`encia), Spain
}%

\date[ Date: ]{\today}
\begin{abstract}

Rotating fermion-boson stars are hypothetical celestial objects that consist of both fermionic and bosonic matter interacting exclusively through gravity. Bosonic fields are believed to arise in certain models of particle physics describing dark matter and could accumulate within neutron stars, modifying some of their properties and gravitational wave emission. Fermion-boson stars have been extensively studied in the static non-rotating case, exploring their combined stability and their gravitational radiation in binary mergers. However, stationary rotating configurations were yet to be found and investigated. The presence of a bosonic component could impact the development of the bar-mode instability in differentially rotating neutron stars. Therefore, the study of rotating fermion-boson stars has important implications for astrophysics, as they could provide a new avenue for the detection of gravitational waves. In addition, these objects may shed light on the behavior of matter under extreme conditions, such as those found in the cores of neutron stars, and explain any tension in the determination of the dense-matter equation of state from multi-messenger observations. In this work we study a new consistent method of constructing uniformly rotating fermion-boson stars and we analyse some of their main properties. These objects might offer alternative explanations for current observations populating the lower black-hole mass gap, as the $2.6 M_\odot$ compact object involved in GW190814.


\end{abstract}
\maketitle

\section{Introduction}

Neutron stars (NSs) are a focal point of numerous studies due to their unique and remarkable properties. They present an exceptional opportunity to examine various areas of physics, including nuclear physics, gravitational physics, and astrophysics. 
Both binary systems and isolated neutron stars are being scrutinized in depth to test General Relativity and different theories of gravity under strong field conditions. Furthermore, with regard to nuclear matter research, recent findings by the Neutron Star Interior
Composition Explorer (NICER) have substantially strengthened constraints on the Equation of State (EOS), and have also furnished highly precise and reliable measurements for the mass and equatorial radius of various millisecond pulsars~\cite{ozel2016measuring,miller2019psr}.

Binary NS systems are considered one of the most promising sources of gravitational waves (GWs)  that can be detected by the advanced LIGO, advanced VIRGO, and KAGRA observatories \cite{akutsu2018kagra,abbott2021searches}. Moreover, the combined use of different types of observations has significantly improved our understanding of compact objects and other astrophysical phenomena. 
We are entering the era of \textit{multi-messenger} astronomy, where the simultaneous detection of signals from multiple messengers (electromagnetic and gravitational waves, neutrinos) results in more accurate and precise measurements of physical quantities associated with NSs~\cite{raaijmakers2020constraining}.

Many NS properties are of significant interest, including quadrupole moments, spin angular velocity, which is directly proportional to the moment of inertia, and their tidal and rotational deformability, often represented by the Love numbers \cite{hinderer2008tidal,postnikov2010tidal}. In August 2017, the first gravitational wave signal from the merger of a compact low-mass NS binary was observed by the LIGO-Virgo detector network~\cite{abbott2017gw170817}, which was followed by an electromagnetic counterpart~\cite{abbott2017gravitational}. Similar systems have since been detected and are currently under investigation~\cite{abbott2020gw190425}. Future new detections of gravitational waveforms emitted by these systems could be used to constrain the parameters of the equation of state of NSs and learn more about their internal composition~\cite{abbott2018gw170817,raithel2019constraints}.

On the other hand, theoretical exotic compact objects could also populated the Universe. One of the simplest examples are boson stars (BSs), compact horizonless, regular solutions of an  ultralight bosonic field theory minimally coupled to gravity, which involves a complex scalar field $\Phi$. However, solutions for massive vector fields (\textit{aka} Proca stars or vector boson stars) have also been obtained, both with and without rotation (for a thorough understanding on BSs, we refer the interested reader to~\cite{Liebling:2012fv,Lai:2004fw,Schunck:2003kk,Brito:2015pxa,herdeiro2019asymptotically,sanchis2019nonlinear}). The concept of BSs dates back to the proposal of \textit{geons} by Wheeler~\cite{PhysRev.97.511} and the introduction of the original spherical scalar BSs by Kaup \cite{PhysRev.172.1331}, Ruffini, Bonazzola, and Pacini \cite{PhysRev.187.1767,PhysRev.148.1269}. The properties of BSs are primarily determined by the potential that encodes the self-interaction of the complex scalar field and varies with the Lagrangian form. Different potentials can be used to model various astrophysical objects, such as black-hole mimickers with features resembling those of NSs or even black holes~\cite{PhysRevD.80.084023,olivares2020tell,herdeiro2021imitation,Pitz:2023ejc,Rosa:2022toh,Rosa:2022tfv,Rosa:2023qcv,Rosa:2024eva,Sengo:2024pwk}. Additionally, BSs and ultralight bosonic fields are considered promising candidates to account for (part of) the dark matter present in galaxy halos \cite{Schunck:1998nq}.

There are several unconventional extensions of BSs, which involve more generalized models of gravity, such as Palatini gravity \cite{Maso-Ferrando:2021ngp}, Einstein-Gauss-Bonnet theory, scalar-tensor models \cite{PhysRevD.56.3478}, or the semi-classical gravity framework \cite{Alcubierre:2022rgp}. Apart from these generalizations, recently considerable effort has been invested into the exploration of more exotic BS models, including multi-state BSs \cite{Urena-Lopez:2010zva}, $\ell$-BSs \cite{Alcubierre:2018ahf}, multi-field multi-frequency BSs~\cite{sanchis2021multifield}, Proca-Higgs stars \cite{Herdeiro:2023lze}, scalar-vector stars, also called \textit{Scalaroca} stars~\cite{Pombo:2023xkw}, and $\ell$-Proca stars~\cite{lazarte2024ell}.

In recent years there has been a surge of interest in the search of additional scalar or vector degrees of freedom. One reason for this was the discovery of the fundamental scalar Higgs boson at CERN \cite{ATLAS:2012yve,CMS:2012qbp}, which has provided a theoretical motivation for the existence of additional scalar fields beyond the Standard Model, such as the axion \cite{PhysRevLett.40.223,PhysRevLett.40.279,Guo:2023hyp} or other ultralight scalar or vector bosons \cite{arvanitaki2010string,Freitas:2021cfi}, which have been proposed as potential dark matter particles \cite{2024arXiv240104735K}. The potential detection of new fundamental vector fields through future observations with LISA is a highly motivating factor for the continued study of these types of objects \cite{Fell:2023mtf}.

As for NSs, GW astronomy presents a new avenue for the exploration of exotic compact objects, which includes BSs. In this auspicious context, an exceptional GW signal was detected in 2020 by advanced LIGO-Virgo, which could potentially be interpreted as the result of a head-on collision of two Proca stars \cite{Bustillo:2020syj} (see also~\cite{bustillo2023searching}). Furthermore, current research efforts are also focused on studying the merger dynamics of BS binaries \cite{palenzuela2017gravitational,Bezares:2019jcb,sanchis2022impact,Bezares:2022obu,siemonsen2023binary,atteneder2024boson}.

If a primordial gas could give rise to bosonic configurations, it is plausible that fermions could also be present during condensation. This suggests the possibility of mixed configurations, consisting of both bosons and fermions. While the original configurations might have been mainly composed of either bosons or fermions, they could also have captured additional fermions and bosons through accretion.

Thus, it is an intriguing theoretical question to explore the properties of these macroscopic composites, which are known as fermion-boson stars in the literature~\cite{Henriques:1989ez}. The initial research on the topic was conducted at the end of $1980s$ \cite{,HENRIQUES198999,Henriques:1990xg,PhysRevD.87.084040}. Other generalizations have also been studied, like charged fermion-boson stars \cite{Kain:2021bwd} and fermion-Proca stars \cite{Jockel:2023rrm}. Numerical simulations have been performed to study their stability and the possible mechanisms through which they could form \cite{valdez2013dynamical,DiGiovanni:2020frc,DiGiovanni:2021vlu,Nyhan:2022pda,DiGiovanni:2022mkn}. It is worth mentioning that recent studies have applied perturbative methods over the static background, obtaining the tidal deformability for mixed stars \cite{Diedrichs:2023trk}.

From an astrophysical point of view, non-zero angular momentum stars are important, and the slowly rotating limit was studied in \cite{deSousa:2000eq}. 
Recent measurements have identified the most rapidly rotating and massive NS in our galaxy to date. Designated as \textit{PSR J0952-0607}, this NS was initially discovered by \cite{Bassa:2017zpe} in 2017. It exhibits an exceptionally short rotational period of $1.41$ milliseconds. Further investigations by \cite{Romani:2022jhd} have revealed that the NS possesses a significantly high mass, estimated to be $M=2.35\pm 0.17 M_{\odot}$. Additionally, the secondary component of the GW event GW190814 could also be a massive NS with $M=2.6M_{\odot}$~\cite{virgo2020gw190814}. These findings provide substantial support for the theory that mixed stars could be plausible explanations for such astrophysical phenomena~\cite{lee2021could,das2021dark,di2022can,valdez2024observations}.
 
In this work we obtain stationary solutions of the Einstein-Euler-Klein-Gordon system, where fermionic and bosonic matter are coupled only through gravity, both have angular momentum and form a spinning mixed fermion-boson star. 

In \Cref{section2}, we establish the theoretical framework, delineating each component and the system as a whole to elucidate the types of objects presented in this manuscript. \Cref{section3} is devoted to the numerical treatment and the development of the algorithm employed for solving the systems. The main global properties of our solutions are obtained in \Cref{section4},  whereas a thorough analysis of the solutions is presented in \Cref{section5}. \Cref{section6} delves into how these types of objects integrate into the GW paradigm, providing some insights into their significance and implications, while \Cref{section7} is devoted to some final remarks and conclusions. Finally, we present the sources of the NSs and mixed stars in Einstein's field equations in~\Cref{appendixA} obtained from the stress energy-tensor, the equations of the initial guess corresponding to the static guess in~\Cref{appendixB}, and further details on the numerical methods in~\Cref{appendixC}. Throughout this work, we adopt $G=c=1$ units.

\section{Formalism }
\label{section2}

We will first introduce the framework to obtain numerical solutions of NSs and BSs, since 
fermionic and bosonic matter describing them behave differently. Besides, the numerical methods for solving each type of star separately are significantly different. In this section, we describe the theoretical set-up and numerical methods to build stationary NSs and BSs separately.

\subsection{Rotating neutron stars}
\label{NSsec}

We can model uniformly rotating NSs described as a perfect fluid assuming a stationary, axisymmetric space-time. The circular velocity  $u^{\alpha}$ can be written in terms of two Killing vectors $t^{\alpha}$ and $\psi^{\alpha}$, being,
\begin{equation}
    u^{\alpha}=u^t(t^{\alpha}+\Omega\psi^{\alpha}),
\label{velocity}    
\end{equation}
where,
\begin{equation}
    u^t\coloneqq\left[g_{\alpha\beta}(t^{\alpha}+\Omega\psi^{\alpha})(t^{\beta}+\Omega\psi^{\beta})\right]^{1/2},
\end{equation}
is the $t$ component of $u^{\alpha}$. In the natural coordinates of $t$ and $\psi$, the angular velocity of the fluid as seen by an observer at rest at infinity can be expressed as follows,
\begin{equation}
    \Omega\equiv \frac{u^{\psi}}{u^t}=\frac{d\psi}{dt}.
\end{equation}
We will have uniform rotation 
if and only if $\Omega=\text{constant}$.
The metric $g_{\alpha\beta}$ for the system can be written as
\begin{equation}
\begin{split}
  ds^2=& -e^{\gamma+\rho}dt^2+e^{\gamma-\rho}r^2\sin^2{\theta}\left(d\psi-\omega dt\right)^2\\
  &+e^{2\alpha} \left( dr^2+r^2d\theta^2\right),
\end{split}
\label{metric1}  
\end{equation}
where $\gamma,\rho,\alpha$ and $\omega$ are functions of $r$ and $\theta$ \cite{friedman2013rotating}. The stress-energy tensor that describes our matter source is given by
\begin{equation}
    T^{\alpha\beta}=\left( \rho_0+\rho_i+ p_{NS}\right)u^{\alpha}u^{\beta}+ p_{NS}g^{\alpha\beta},
    \label{tmnfermion}
\end{equation}
where $\rho_0$ is the rest-energy density, $\rho_i$ is the internal energy density and $p_{NS}$ is the pressure. The Einstein field equations read

\begin{equation}
    G_{\alpha\beta}=R_{\alpha\beta}-\frac{1}{2}Rg_{\alpha\beta}=8\pi T_{\alpha\beta},
\end{equation}
where $R_{\alpha\beta}$ and $R$ are the Ricci tensor and scalar.
We solve the field equations coupled to the Euler equations in order to obtain solutions of rotating NSs.
 To do this we will employ the RNS code \cite{stergioulas1994comparing}, which is based on the Komatsu, Eriguchi, and Hachisu (KEH) method , also known as the Hachisu self-consistent field (HSCF) \cite{komatsu1989rapidly}. This method makes use of a Green's function approach, in which the Einstein field equations are manipulated to isolate terms for which the flat-space Green's function is known. The remaining terms (referred to as the sources) are moved to the right-hand side. The metric functions are then determined at each iteration through an integral involving the sources (including the stress-energy tensor and other elements) and Green's functions. This approach combines different aspects of spectral methods and finite difference techniques, and was modified by Teukolsky, Cook, and Shapiro \cite{saphiro}.

Taking the above into account, the Einstein field equations will be presented as second-order partial derivative operators over some combinations of the metric potentials, equated to the source terms,
\begin{eqnarray}
    \Delta\left[\rho e^{\gamma/2}\right]=S_{\rho}(r,\mu),
    \label{equationrho}\\
    \left(  \Delta +\frac{1}{r}\partial_r-\frac{\mu}{r^2}\partial_{\mu}\right)\left[\gamma e^{\gamma/2}\right]=S_{\gamma}(r,\mu),
    \label{equationgama}\\
    \left(  \Delta +\frac{2}{r}\partial_r-\frac{2\mu}{r^2}\partial_{\mu}\right)\left[\omega e^{(\gamma-\rho)/2}\right]=S_{\omega}(r,\mu),
    \label{equationomega}
\end{eqnarray}
with $\Delta$ the flat-space, spherical coordinate Laplacian, $\mu=\cos\theta$, and $S_{\rho}$, $S_{\gamma}$, $S_{\omega}$ are the mentioned sources, shown in the Appendix \ref{appendixA}. The fourth field equation determines  $\alpha$ and it is not separable in this way.
The four field equations together with the hydrostatic equilibrium one,
\begin{equation}
dp_{NS}-\left(\rho_0+\rho_i+p_{NS}\right)\left[d\ln u^t+u^tu_{\psi}d\Omega\right]=0,
  \label{hydrostaticeq}  
\end{equation}
and a given EOS closes the problem.
The above approach guarantees that the selected metric functions exhibit asymptotically flat behavior at infinity. For the three differential metric equations detailed in \cref{equationrho,equationgama,equationomega}, the boundary conditions are inherently met at spatial infinity by design \cite{friedman2013rotating}.
The behavior of the fermionic matter in this work is assumed to follow a polytropic law for simplicity,
\begin{equation}
    p_{NS}=K\rho_{NS}^{1+1/n},
\end{equation}
where $\Gamma=1+1/n$ is the adiabatic index and $K$ is the polytropic constant, which will determine our scale all over the work. Further, $\rho_{NS} = \rho_0 + \rho_i$ is the total energy density. Since $K^{n/2}$ has units of length, we can use it to redefine dimensionless quantities and fix the scale of our system.

We will define the zero angular momentum observer (ZAMO) velocity,
\begin{equation}
    v=(\Omega-\omega)r\sin\theta e^{-\rho},
\end{equation}
which will be useful in what follows.

The metric functions are numerically determined by integrating the sources shown in \Cref{appendixA}, where we use the method of the Green functions for the radial part, and the corresponding polynomials for the angular dependence. This is the RNS philosophy.

\subsection{Boson stars}

 The complex scalar field dynamics are described by the Lagrangian,
\begin{equation}
 \mathcal{L}_{\Phi}=-\frac{1}{2}\left[g^{\alpha\beta}\nabla_{\alpha}\Phi^*\nabla_{\beta}\Phi+V\left(|\Phi|^2\right)\right],
    \label{lagrangian}
\end{equation}
where $V\left(|\Phi|^2\right)$ is a potential that depends only on the absolute value of the scalar field, respecting the global $U(1)$ invariance of the model. 

Through a minimal coupling to gravity, the above enters the action in the following manner
\cite{Liebling:2012fv},
\begin{equation}
    \mathcal{
    S}=\int \left(\frac{1}{16\pi }R+\mathcal{L}_{\Phi}\right)\sqrt{-g}d^4x.
    \label{action}
\end{equation}

This equation, called the Einstein-Klein-Gordon (EKG) action, describes the behavior of a massive complex scalar field $\Phi$  coupled to Einstein's gravity. In this equation, $g$ is the metric determinant.

Varying the action (\ref{action}) yields the EKG equations,
\begin{equation}
\begin{split}
    &R_{\alpha\beta}-\frac{1}{2}Rg_{\alpha\beta}=8\pi T_{\alpha\beta}, \\
    &
    g^{\alpha\beta}\nabla_\alpha\nabla_{\beta}\Phi=\frac{dV}{d|\Phi|^2}\Phi,
\label{kg}
\end{split}
\end{equation}
where  $T_{\alpha\beta}$ is now the canonical stress-energy tensor of the scalar field,
\begin{equation}
\begin{split}
 T_{\alpha\beta}=&\frac{1}{2}\left(\nabla_{\alpha}\Phi^*\nabla_{\beta}\Phi+\nabla_{\beta}\Phi\nabla_{\alpha}\Phi^*\right)-\\
 &\frac{1}{2}g_{\alpha\beta}\left[g^{\mu\nu}\nabla_{\mu}\Phi^*\nabla_{\nu}\Phi+V\left(|\Phi|^2\right)\right].
    \label{stress}
\end{split}
\end{equation}
 Equilibrium solutions are obtained
assuming an axisymmetric coordinate system as in~\cref{metric1} and the following ansatz for the field, as in \cite{Herdeiro:2015gia,PhysRevD.55.6081}, 
\begin{equation}
     \Phi(t,r,\theta,\psi)=\phi(r,\theta)e^{-i(w t+\eta\psi)}.
     \label{scalar}
 \end{equation}
 The angular frequency of the field is $w \in \mathbb{R}$ 
 and $\eta \in \mathbb{Z}$ is the harmonic index, also called the winding number and the complex scalar field modulus is $| \Phi(t,r,\theta,\psi)|=\phi(r,\theta)$.

\subsection{Mixed stars and rotating fermion-boson stars}
\label{mixedrotstars}

Several studies have been conducted on the static case of mixed fermion-boson configurations, also called fermion-boson stars and neutron-boson stars (NBSs)  \cite{DiGiovanni:2020frc,Nyhan:2022pda,DiGiovanni:2022mkn,DiGiovanni:2021vlu,PhysRevD.87.084040,Kain:2021bwd}. However, since rotation is a ubiquitous phenomenon in astrophysics, we aimed to develop a framework for solving the Einstein equations for mixed rotating stars, which are more closely aligned with plausible astrophysical scenarios.

In this initial approach, we have developed a code that can solve the Einstein equations for an axisymmetric spacetime with a matter content of fermions and bosons that are coupled solely by gravity, 
\begin{equation}
    R_{\alpha\beta}-\frac{1}{2}Rg_{\alpha\beta}=G_{\alpha\beta}=8\pi (T^f_{\alpha\beta}+T^b_{\alpha\beta}).
    \label{fulleinstein}
\end{equation}
The metric ansatz is precisely \cref{metric1}. The stress-energy tensor for the fermionic matter, i.e. the neutron star component, is represented by $T_{\alpha\beta}^{f}$ and it is exactly \cref{tmnfermion}, while $T_{\alpha\beta}^{b}$ represents the stress-energy tensor for the bosonic matter. The Einstein equations can be reformulated in terms of the aforementioned differential operators and sources,
\begin{equation}
    \begin{split}
    &\Delta\left[\rho e^{\gamma/2}\right]=S_{\rho}^T(r,\mu),\\
    &
    \left(  \Delta +\frac{1}{r}\partial_r-\frac{\mu}{r^2}\partial_{\mu}\right)\left[\gamma e^{\gamma/2}\right]=S_{\gamma}^T(r,\mu),
    \\
    &
    \left(  \Delta +\frac{2}{r}\partial_r-\frac{2\mu}{r^2}\partial_{\mu}\right)\left[\omega e^{(\gamma-\rho)/2}\right]=S_{\omega}^T(r,\mu),
    \end{split}
    \label{equationsourcestot}
\end{equation}
The sources corresponding to the bosonic part are obtained using the high coupling-constant or Ryan-Pani-Vaglio approximation, following the seminal authors' works \cite{Ryan:1996nk,Vaglio:2022flq} where such approach is presented, explained, and applied. In what follows, we use both terminologies indistinctly to refer to this framework. The justification for the use of such an approach is based on both stability and simplicity. First, it has been shown that for spinning BS, non-axisymmetric instabilities (known in the literature as NAI \cite{DiGiovanni:2020ror,Dmitriev:2021utv,Siemonsen:2020hcg}), can be avoided by introducing the quartic self-interaction potential used in this approach. Specifically, high values of the coupling constant (as it is our case) quench the non-axisymmetric instabilities of spinning BSs. Stable rotating BSs would then be described by a quartic or higher self-interacting potential with a large value of the coupling constant, justifying the use of such a framework in our study. Also, it has been shown that mixed stars with different self-interacting potentials could also be stabilized due to the presence of the fermionic matter in the system~\cite{DiGiovanni:2021vlu}. Second, the stress energy tensor takes the form of a perfect fluid in the high-coupling-constant limit. The implementation of a numerical approach considering potentials under which the stress-energy tensor can not be taken as a perfect fluid would be considerably more complicated.

It is important to notice that when computing the source terms for the mixed star, we have contributions from the fermionic and bosonic matter but also some curvature contributions that cannot be computed independently. That is to say, with a total stress-energy tensor given by $T^T_{\alpha\beta}=(T^f_{\alpha\beta}+T^b_{\alpha\beta})$, the full sources $S_i^T\neq S_i^f+S_i^b$.

In the above-mentioned framework, we have the following potential:
\begin{equation}
    V(|\Phi|)=\mu_b^2|\Phi|^2+\frac{1}{2}\lambda |\Phi|^4, 
\end{equation}
where $\mu_b$ is the boson scalar field mass (not to be confused with $\mu=\cos\theta$) and $\lambda$ the coupling or self-interacting constant. In the high-coupling approximation, although the ansatz for the field is \cref{scalar}, we marginalize the bosonic tail and assume that the scalar field in the \textit{outer region} is $|\Phi|=0$, while, in the inner part,
\begin{equation}
    |\Phi|^2={\rm{max}}\left[0, \frac{1}{\lambda}\left(\frac{(w-\eta\omega)^2}{e^{\gamma+\rho}}-\frac{e^{\rho-\gamma}\eta^2}{r^2\sin^2\theta}-\mu_b^2\right)\right] .
\end{equation}
Moreover, the stress-energy tensor for the bosonic matter reduces to that of a perfect fluid describing the interior of the star, while the exponential tail region identified as the \textit{outer part} is neglected under this approximation, 
\begin{equation}
    T_{\mu\nu}^{b}=(\rho_{BS}+p_{BS})\hat{u}_{\mu}\hat{u}_{\nu}+p_{BS}g_{\mu\nu}.
\end{equation}
The pressure and energy density are given by
\begin{equation}
    p_{BS}=\frac{1}{4}\lambda |\Phi|^4,\hspace{0.5cm}\rho_{BS}=\mu_b^2|\Phi|^2+\frac{3}{4}\lambda |\Phi|^4,
\end{equation}
 and the four-velocity,
 \begin{equation}
     (\hat{u}_t,\hat{u}_r,\hat{u}_{\theta},\hat{u}_{\psi})=\frac{(-w,0,0,\eta)}{(\lambda |\Phi|^2+\mu_b^2)^{\frac{1}{2}}} .
 \end{equation}
So the proper velocity of the zero angular momentum observer is given by,
\begin{equation}
    \Bar{v}=\frac{\eta}{(w-\eta\omega)}\frac{e^{\rho}}{r\sin\theta} .
\end{equation}
As under this approximation the stress-energy tensor can be re-expressed in a perfect-fluid-like form, we obtain the source equations analogously to the NS case. This approach can also be called the Ryan-Pani-Vaglio approximation. We also have defined the energy density and pressure induced by the bosonic part. Therefore, the equations for the mixed stars will be similar to the NS ones, but with additional bosonic matter terms. The corresponding sources are shown in Appendix \ref{appendixA}. As discussed in \Cref{NSsec}, the regularity and asymptotic flatness of the metric functions, specifically the boundary conditions, are inherently ensured by the construction of the KEH method \cite{friedman2013rotating}.

\section{Numerical implementation }
\label{section3}
\subsection{Internal scaling}
\label{section3a}

Our approach consists in using the RNS as the base code, and implementing the Ryan-Pani-Vaglio approximation for the BSs as additional source terms in the usual NSs algorithm. The code had to be modified to include the bosonic matter part but our algorithm is essentially the same RNS code,  
making use of the HSCF method based on Green functions to solve our system of equations. The main problem we dealt with was the different rescaling for both types of objects. As shown in the RNS bibliography~\cite{saphiro}, the whole code uses a map between the compactified coordinate $s$ and the full spherical radius $r$ through the equatorial coordinate radius $r_e$, following
\begin{equation}
r = r_e\left(\frac{s}{1-s}\right).
\end{equation}
%

With the associated differentiation rule shown in Appendix \ref{appendixB}. For the NS case, it is straightforward to obtain $r_e$, since it is the coordinate radius at the star's surface, but when dealing with BSs or mixed stars, in the static case, it is not clear how to define this quantity, as the bosonic matter decays exponentially to zero at infinity in a non-compact form. This parameter is important because the method uses this radius as a scaling parameter. For each star, we should have a different value of $r_e$. To solve this issue, we have implemented the following criterion: on one hand, in the pure BS case, the $r_e$ is defined by the radius at which $99\%$ of the scalar field energy density is enclosed by a sphere of that radius. On the other hand, when we have a mixed star as the initial guess we simply take the fermionic $r_e$.
We have checked the consistency and stability of the method for various mixed and purely bosonic spherical static stars, used as initial guesses, by having close but different values for $r_e$.

In this case, where the initial guesses are only slightly different, the rotating solutions obtained for a given set of model parameters are identical. This indicates that the code converges to the same solution when given similar initial conditions for specific parameters in the rotating scenario. The calculations involving the initial guesses can be found in \cref{section3B} and are explicitly shown in Appendix \ref{appendixB}.


As mentioned before, the scaling for the problem is defined by the polytropic constant $K$, i.e, the NS part. From~\cite{Cook:1993qj} we have the following dimensionless re-definitions,
\begin{equation}
    \begin{split}
        &\Bar{r}\equiv K^{-n/2}r,\\
        &\Bar{t}\equiv K^{-n/2}t,\\
        &\Bar{\omega}\equiv K^{n/2} \omega,\\
        &\Bar{\Omega}\equiv K^{n/2} \Omega,\\
        &\Bar{\rho}_0\equiv K^n \rho_0,\\
        &\Bar{\rho}_i\equiv K^n \rho_i,\\
        &\Bar{p}_{NS}\equiv K^n p_{NS}.
    \end{split}
\end{equation}
For the BS part, the scaling with the polytropic constant may seem less natural than using the boson mass $\mu_b$, but as it is easier to use same scaling to build mixed configurations, we will maintain both the self-interaction constant and the mass parameter. The re-scaled and dimensionless BS parameters are given by
 \begin{equation}
    \begin{split}
        &\Bar{w}\equiv K^{n/2}w,\\
        &\Bar{\eta}\equiv K^{-n} \eta,\\
        &\Bar{\mu_b}\equiv K^{-n/2}\mu_b,\\
        &\Bar{\lambda}\equiv  K^{-n} \lambda.
    \end{split}
\end{equation}
This also leads to the usual constraint for the frequency defined by $\Bar{w}\in [0,\Bar{\mu_b}]$.
It is important to note that, at the level of the code, we do an effective re-scaling by fixing $\mu_b$ to unity, but as this boson mass is also re-scaled by our choice of the natural parameter for the problem (the polytropic constant), we cannot just remove it from the code, and it will appear as $\Bar{\mu_b}$. 

Having the aforementioned re-scaling rule settled, the observable quantities obtained are also dimensionless, and the transformation law is defined by,
\begin{equation*}
    \begin{split}
        &\Bar{M}=K^{-n/2}M,\\
        &\Bar{J}=K^{-n}J,
    \end{split}
\end{equation*}
with $M$ and $J$ the mass and the angular momentum of the star, respectively.

\subsection{ Initial guess, shooting method, and coordinate transformation}
\label{section3B}

In principle, there exist three possibilities for an initial guess, namely the purely fermionic (or NS), the purely bosonic, and the mixed case. In the present paper we focus on rotating stars with a non-negligible fermionic component, where purely fermionic or mixed initial guesses are more appropriate. When we consider a larger fermionic component we refer to the ratio between the fermion to the total mass. Concretely, the numerical results which we display explicitly in the subsequent sections were obtained from purely fermionic initial guesses, because the required numerical calculations are much faster than for the mixed case. We want to emphasize, however, that we performed a significant number of calculations for mixed initial guesses, to make sure that the resulting rotating stars do not depend on the initial guesses. The procedure of how to take a spherically symmetric star as an initial guess for a spinning star is well explained in the RNS manual \cite{stergioulas1992rotating}. 
For completeness, we will nevertheless briefly explain how to implement all possible cases of initial guesses: the static NS, the static BS, and the static mixed star.  



For the pure NS case as initial guess, the Tolman-Oppenheimer-Volkoff (TOV) equations are solved for the given polytropic parameters, with the line element for a static spherical object defined by
\begin{equation}
    ds_{SNS}^2=-e^{2\nu(r)}dt^2+\frac{1}{1-2m(r)/r}(dr^2+r^2d\Omega),
\end{equation}
where (SNS) means static NS.

As in the RNS code, we introduce the auxiliary variable $\Upsilon$ defined by
$e^{2\Upsilon(r)}=\frac{1}{1-2m(r)/r}$, which simplifies the \textit{Radial Einstein equation} (REe). 
Thus, we have static NSs described by the following system of equations:

\begin{equation}
\begin{split}
&\text{EOS:} \hspace{2.15cm} p_{NS}=p_{NS}(\rho_{NS})  \\
&\text{Mass:}\hspace{2cm}  \frac{dm(r)}{dr}=4\pi r^2 \rho_{NS}(r)  \\
&\text{REe:} \hspace{0.4cm} \frac{d\nu(r)}{dr}=\frac{8\pi re^{2\Upsilon(r)} p_{NS}(r)-\Upsilon'(r)\left(2+r\Upsilon'(r)\right)}{2+2r\Upsilon'(r)} 
\\
&\text{TOV:} \hspace{2cm}\frac{dp_{NS}(r)}{dr}=-\nu'(r)(\rho_{NS}(r)+p_{NS}(r))  ,
\end{split}
\label{constitutivas1}
\end{equation}
where now $m'$ in the \textit{Mass equation} must be substituted by $m'(r)=\frac{1}{2}e^{-2\Upsilon(r)}\left(e^{2\Upsilon(r)}+2r\Upsilon'(r)-1\right)$ .

Both the original RNS code and our extension use a Runge-Kutta method to integrate Equations~(\ref{constitutivas1}). Next, a subroutine interpolates the interior of the stars and matches them with the outer solution. This yields the metric functions in a suitable form for their usage as initial guesses in the rotation extension. The static star employed as the initial guess for the metric \cref{metric1} results in the following isotropic coordinates
\begin{eqnarray}
\mbox{INT:} && ds^2=-e^{2\nu}dt^2+e^{2\xi}(dr^2+r^2d\theta+r^2\sin^2\theta d\psi^2) \nonumber \\
\Rightarrow && \gamma=\nu+\xi,\hspace{0.2cm}\rho=\nu-\xi,\hspace{0.2cm}\alpha=(\gamma-\rho)/2,\hspace{0.2cm}\omega=0 \nonumber \\
\mbox{OUT:} && e^{(\gamma+\rho)/2}=[1-M_s/(2r)]/[1+M_s/(2r)], \nonumber \\
\Rightarrow &&e^{(\gamma-\rho)/2}=e^{\alpha}=(1+M_s/(2r))^2,\hspace{0.2cm}\omega=0.
\label{constitutivas2}
\end{eqnarray}

Where $M_s$ is the gravitational mass of the star~\cite{Weinberg:1972kfs}.
Through the previous transformations, the initial guess calculation for the scenario where the initial guess is purely fermionic , is straightforward.

Formulating an initial guess for a mixed or purely bosonic star requires a slightly different approach compared to the NS case.
The mixed and purely bosonic scenario can be obtained with a single, general formalism, by choosing a zero or non-zero fermionic energy density at the origin for pure BSs or mixed stars. It is important to note that the Ryan-Pani-Vaglio approximation is not required within a static framework. The static case can be addressed using a one-parameter, one-dimensional shooting method for a configuration with a non-zero field at the origin. In this section, we will provide a detailed explanation of the algorithm.

In contrast to the pure NS case, where the RNS code assumes a metric in isotropic coordinates, the metric for the mixed and bosonic cases must be expressed in general, static, spherically symmetrical Schwartzschild-like coordinates, and the TOV system needs the corresponding modifications. Our chosen metric in these cases is the same as the one used in \cite{DiGiovanni:2020frc},
\begin{equation}
    ds^2=-b^2dt^2+a^2dr^2+(r^2d\theta+r^2\sin^2\theta d\psi^2) .
    \label{metricbs}
\end{equation}
The scalar field ansatz in spherical symmetry is now 
\begin{equation}
    \Phi_s(t,r)=\phi_s(r)e^{-iw_st}.
\end{equation}
The modified TOV system is
\begin{widetext}
\begin{equation}
    \begin{split}
        &\frac{db}{dr}=\frac{b}{2}\left(\frac{a^2-1}{r}+8\pi r\left[\left(\frac{w_s^2}{b^2}-\mu_b^2-\frac{\lambda}{2}\phi_s^2\right)a^2\phi_s^2+\Psi^2+2a^2p_{NS}\right]\right),\\
         &\frac{da}{dr}=\frac{a}{2}\left(\frac{1-a^2}{r}+8\pi r\left[\left(\frac{w_s^2}{b^2}+\mu_b^2+\frac{\lambda}{2}\phi_s^2\right)a^2\phi_s^2+\Psi^2+2a^2\rho_{NS}\right]\right),\\
         &\frac{d\phi_s}{dr}=\Psi,\\
         &\frac{d\Psi}{dr}=-\left(1+a^2-8\pi r a^2\left[\mu_b^2\phi_s^2+\frac{\lambda}{2}\phi_s^4+\frac{1}{2}(\rho_{NS}-p_{NS})\right]\right)\frac{\Psi}{r}
         -\left(\frac{w_s^2}{b^2}-\mu_b^2-\lambda\phi_s^2\right)a^2\phi_s,\\
         &\frac{dp_{NS}}{dr}=-(\rho_{NS}+p_{NS})\frac{1}{a}\frac{db}{dr}.
    \end{split}\label{syseq}
\end{equation}
\end{widetext}
Together with the above system of equations~(\ref{syseq}), we give the equation for the gravitational mass,
\begin{equation}
M_s=\lim_{r\rightarrow\infty}\frac{r}{2}\left(1-\frac{1}{a^2}\right).
\end{equation}
For clarity, we have omitted the functional dependencies.
This modified TOV system allows us to recover the static pure BS when $p_{NS}=\rho_{NS}=0$.
This set of equations defines an eigenvalue problem and it can be solved, as mentioned previously, using a shooting method over the frequency parameter. For a given central field value $\phi_0=\phi_s(r)|_{r=0}$,  there is a $w_{shot}$ that ensures the right behavior at the boundaries. The differential equations are solved using the shooting method for a Runge-Kutta $4th$ order.
After finding the correct eigenvalue, we have to rescale both the frequency and $g^{tt}$ by its value at the outer boundary $g^{tt}(r\rightarrow \infty)$.

After solving the system with the shooting, we need a metric transformation to set the initial guess as our code requires. The transformation might allow the code to read the solved static mixed or pure BS obtained from \cref{metricbs} in terms of the axisymmetric language \cref{metric1}.  A coordinate transformation is mandatory, and for the static scenario,  the data must be reformulated in such a way that the metric functions derived for the static star are analogous to those of the rotating system in the limit where 
$\omega=0$. For clarity, we show the calculations in detail in Appendix \ref{appendixB} .

\subsection{Green's functions and integration}
\label{integration_method}
As mentioned in \cref{NSsec}, the RNS code \cite{saphiro} and our modified code exploit the Komatsu, Eriguchi, and Hachisu self-consistent field method \cite{komatsu1989rapidly}, in turn inspired by the seminal works presented by Butterworth and Ipser \cite{1976ApJ...204..200B}. We will explain in detail how the integration using Green's functions is performed. The field equations in the operational form in \cref{equationsourcestot} are elliptical, and taking into account the equatorial and axial symmetry of the problem, the metric potentials can be written as integral forms in terms of three-dimensional Green's functions. The main idea underlying this method is precisely based on this integral representation of the field equations for each independent component of the metric functions. Using appropriately the corresponding Green's functions in each case, the static solution is taken as the initial guess, which is iterated until the field equations are satisfied up to a chosen tolerance. This concept can be elucidated as follows: Einstein's equations are divided such that, on one side of the equality, we find the Green's functions in flat space, which are well-known and readily solvable. On the opposite side are the remaining source terms. During each iteration, the metric functions are updated by integrating the product of these Green's functions with the source terms \cite{Ryan:1996nk}.


The three elliptic potentials $\rho$, $\gamma$, and $\omega$ are derived by expanding the Green functions in their radial and angular parts (as explained in \cite{komatsu1989rapidly} and described in Appendix \ref{appendixC} for one particular case). These potentials are detailed in Appendix \ref{appendixC}. The source terms $\widetilde{S}_{\rho}^T(s,\mu)=\bar{r}^2S_{\rho}^T(s,\mu)$, $\widetilde{S}_{\gamma}^T(s,\mu)=\bar{r}^2S_{\gamma}^T(s,\mu)$, and $\widetilde{S}_{\hat{\omega}}^T(s,\mu)=\bar{r}_e\bar{r}^2S_{\omega}^T(s,\mu)$ are integrated using the Legendre and associated polynomials $P_n(\mu)$ and $P_n^m(\mu)$, with trigonometric functions expressed in terms of $\mu$ through their inverse functions, e.g., $\sin(n\theta) = \sin(n\arccos\mu)$.
The method ensures the correct asymptotic behavior of the elliptic potentials, where for large $\Bar{r}$, $\rho\sim\mathcal{O}\left(1/\Bar{r}\right)$,$\gamma\sim\mathcal{O}\left(1/\Bar{r}^2\right)$, and $\omega\sim\mathcal{O}\left(1/\Bar{r}^3\right)$. Appendix \ref{appendixC} provides the identities used to improve accuracy.
 It is crucial to note that the equation for $\alpha$ \ref{alpa}, corresponding to the fourth field equation, is not an elliptic equation but rather an ordinary first-order differential equation. This can be derived by integrating the derivative expression detailed in \Cref{appendixA}, together with the condition at the pole $\alpha=\frac{1}{2}(\gamma-\rho)$ \cite{Vaglio:2022flq}.

\vspace{0.5cm}

It is imperative to acknowledge the significance of Ridder's method within our computational framework. The second version of the RNS code incorporates this root-finding algorithm specifically to generate sequences of polytropic NS solutions, as outlined in \cite{stergioulas1994comparing}. In a parallel methodology, we have adopted Ridder's method with the analogous aim of computing sequences for our mixed systems, now using $\Bar{w}$ as the sequence parameter.

This extra algorithm is initiated with a given value for the field frequency $\Bar{w}$, progressively reducing it by a fixed amount until reaching a determined value. We then input the oblateness through an initial value for $r_{ratio}=r_p/r_e$, which is the fraction of polar over the equatorial radii. A value $r_{ratio}=1$ corresponds to the spherical and static case, while $r_{ratio}<1$  includes rotation in the system. For each $\Bar{w}$ value in the loop, there is a true solution for a fixed $r_{ratio}$. Ridder's method is able to find this solution for the initial conditions -- given as the input and jumps to the next $\Bar{w}$ -- when the difference between the value for $r_{ratio}$ and $r_{ratio}+\delta r_{ratio}$ is less than a fixed tolerance, being $\delta r_{ratio}<<r_{ratio}$.

A comprehensive and detailed exposition of Ridder's method is provided in \cite{PresTeukVettFlan92}, offering crucial insights into its application and efficacy.

\section{Global properties}
\label{section4}

Once the equations for the metric functions are solved, the main observable quantities can be extracted in terms of the Killing vectors $\xi^{\beta}_{(t)},\xi^{\beta}_{(\psi)}$ and the stress-energy tensor using the Komar integrals \cite{stergioulas1994comparing}
\begin{equation}
\begin{split}
    & M=-\int(2 T^{\alpha}_{\beta}-\delta^{\alpha}_{\beta}T^{\sigma}_{\sigma})\xi^{\beta}_{(t)}d^3\Sigma_{\alpha}=\int (-2T^t_{t}+T^{\sigma}_{\sigma})\sqrt{-g}d^3x.\\
    &J=\int T^{\alpha}_{\beta}\xi^{\beta}_{(\psi)}d^3\Sigma_{\alpha}=\int T^t_{\psi}\sqrt{-g}d^3x.
    \end{split}
\end{equation}

Since we have $T^T_{\alpha\beta}=(T^f_{\alpha\beta}+T^b_{\alpha\beta})$ we can obtain both integrals separately and, through a simple addition, we will have the total mass and angular momentum. The mass and angular momentum for the NS part is given by \cite{saphiro} as follows,
 \begin{eqnarray}
         \bar{M}_{NS}&=&
         4\pi \bar{r}_e^3\int_0^1\frac{s^2ds}{(1-s)^4}\int_0^1d\mu e^{2\alpha+\gamma}\left[\frac{(\bar{p}_{NS}+\bar{\rho}_0+\bar{\rho}_i)}{1-v^2}\times\right.\nonumber\\
         &&\left.\left(1+v^2 +\frac{2vs}{1-s}\sqrt{1-\mu^2}e^{-\rho }\Bar{\omega}\right) +2\bar{p}_{NS}\right],\\
        \bar{J}_{NS}&=&4\pi \bar{r}_e^4\int_0^1\frac{s^3ds}{(1-s)^5}\int_0^1d\mu \sqrt{1-\mu^2} e^{2\alpha+\gamma-\rho}\times\nonumber\\
        &&\frac{\displaystyle(\bar{p}_{NS}+\bar{\rho}_0+\bar{\rho}_i)v}{\displaystyle 1-v^2}.
\end{eqnarray}

 And the BS part, 
 \begin{eqnarray}
      \Bar{M}_{BS}&=&4\pi \bar{r}_e\int_0^1\frac{ds}{(1-s)^2}\int_0^1d\mu e^{2\alpha-\rho}|\Phi|^2\left[e^{2\rho}\frac{\bar{\eta}^2}{1-\mu^2}+\right.\nonumber\\
      &&\left.\frac{s^2}{(1-s)^2}\left(\bar{w}^2-\bar{\eta}^2\bar{\omega}^2+e^{\gamma+\rho}\bar{\lambda}\frac{|\Phi|^2}{2}\right)\right] ,\\ 
     \Bar{J}_{BS}&=&4\pi \bar{r}_e^3\int_0^1\frac{s^2ds}{(1-s)^4}\int_0^1d\mu e^{2\alpha-\rho}\bar{\eta}(\bar{w}-\bar{\eta}\bar{\omega})|\Phi|^2.
\end{eqnarray}

We keep both quantities separate and treat them as completely independent observables.,
 \begin{equation}
     \begin{split}   &\bar{M}_{T}=\bar{M}_{NS}+\bar{M}_{BS},\\  &\bar{J}_{T}=\bar{J}_{NS}+\bar{J}_{BS}.
     \end{split}
 \end{equation}
There is another method for obtaining these quantities, making use of the relations between the sources and the observables, just by integrating with the proper Legendre polynomials. We have the general formulae for the mass and current moments \cite{Vaglio:2022flq},
\begin{eqnarray}
    M_{2n}&=&\frac{1}{2}r_e^{2n+1}\int_0^1 ds \frac{s^{2n}}{(1-s)^{2n+2}}\int_0^1 d\mu' P_{2n}(\mu')\nonumber\\
    &&\times S_{\rho}(s,\mu'),\\
    S_{2n-1}&=&\frac{r_e^{2n+1}}{4n}\int_0^1  \frac{dss^{2n}}{(1-s)^{2n+2}}\times\nonumber\\
    &&\int_0^1d\mu'\sin\theta'P_{2n-1}^1(\mu')S_{\omega}(s,\mu'),
\end{eqnarray}
with $n$ the multipole order, and $P_{m}^l$ the family of Legendre polynomials and their derivatives. We can directly integrate the expression for $M_0$ which is the mass, $S_1$ for the angular momentum, and beyond for the quadrupole moment and higher order multipole moments. Note that our expressions for the multipole integrals are slightly different from the ones in \cite{Vaglio:2022flq} since our sources, at the level of the code, are re-scaled with a factor of $\bar{r}^2$. The above integrals are key to perform checks on the calculations of our code and to obtain higher order multipoles in future works.


\section{Equilibrium configurations}
\label{section5}
\subsection{Static solutions}
\label{sol_stat}
\begin{figure}[]
\includegraphics[clip,width=1.0\columnwidth]{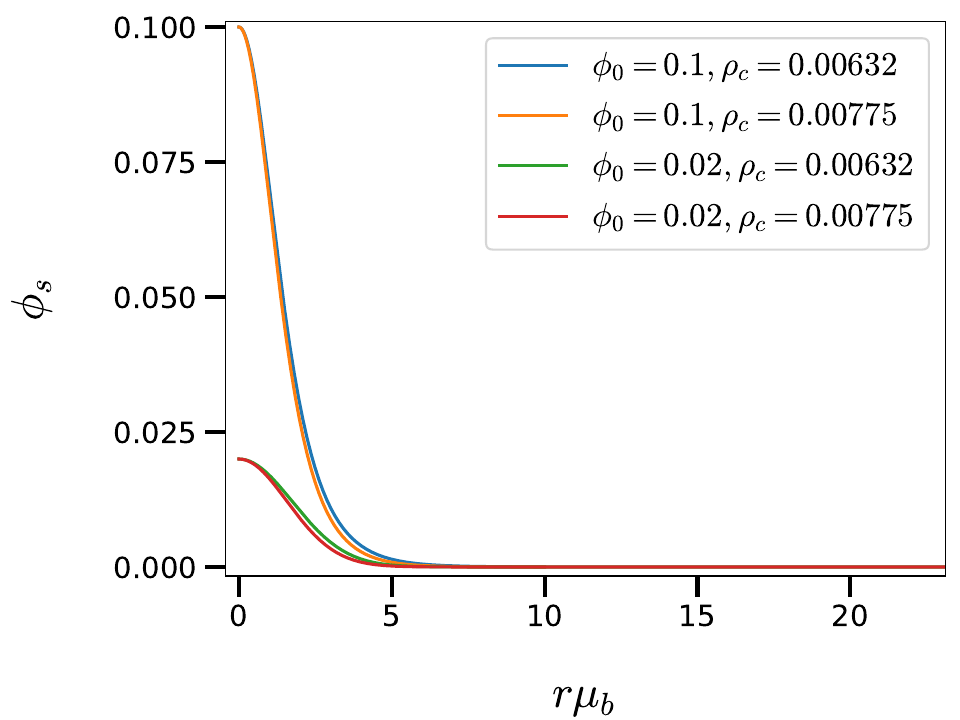}\\%
\includegraphics[clip,width=1.0\columnwidth]{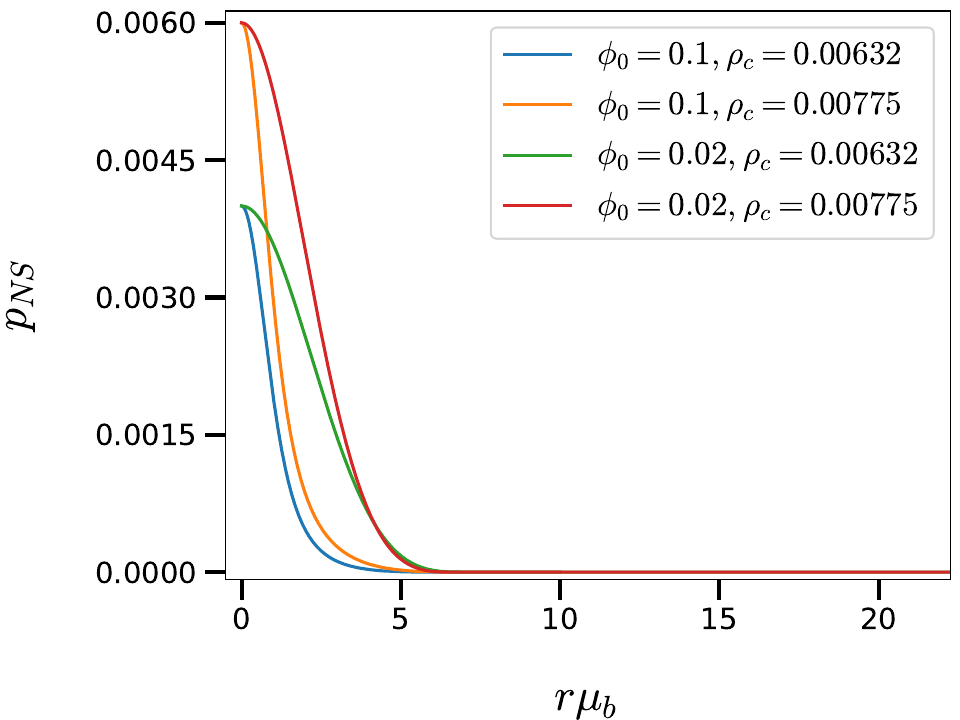}%
\caption{Radial profiles of the scalar field $\phi_s$ (top panel) and fermionic pressure $p_{NS}$ (bottom panel) for four different static mixed star models. The models have a total mass of  $M_s\mu_b=\lbrace0.88,1.172,1.922,1.887\rbrace$. 
The value of the coupling constant is $\Lambda=100$.}
\label{static}
\end{figure}

The investigation of static solutions is of some independent interest, but they also serve as initial guesses for spinning solutions, although in this work we mainly used static pure fermion stars as initial guesses, as explained at the beginning of \cref{section3B}. Examining the primary characteristics of these static stars allows us to perform comparative analyses with previous studies and identify some similarities and differences with spinning configurations. Our findings are consistent with the solutions presented in~\cite{Valdez-Alvarado:2012rct,Valdez-Alvarado:2020vqa,DiGiovanni:2020frc,DiGiovanni:2021vlu}, although some variations arise due to different self-interaction coupling constant values in our setup, reflecting the high-value nature of the coupling in our framework.
It is also important to notice that the re-scaling for the static case is not the same as in the spinning case, and as mentioned in \cref{section3B}, no approximations are used. Now we maintain the polytropic constant value intact, and the re-scaling is done in terms of the boson mass term $\mu_b$, having $r\rightarrow r\mu_b$, $t\rightarrow t\mu_b$, $w\rightarrow w/\mu_b$ and $M_s\rightarrow M_s\mu_b$. Also, the coupling constant is redefined as $\lambda\rightarrow \lambda/(4\pi \mu_b^2)=\Lambda$, which is the usual rescaling for scalar boson stars \cite{Liebling:2012fv,DiGiovanni:2021ejn}. This is the reason why we do not have bar quantities for the static case.
The bosonic field is depicted in the upper panel of \cref{static}, while the lower panel illustrates the fermion pressure as a function of radius for four distinct models of mixed stars.  

The usual behaviour in static mixed stars can be seen in \cref{static}. First, the field going to zero at infinity. Second, the polytropic fermion pressure also has a continuous behavior at the NS surface. The obtained masses are within the usual range of values for these objects, i.e $M_s\mu_b\in [0.5,2)$ \cite{Valdez-Alvarado:2012rct,DiGiovanni:2020frc,DiGiovanni:2021vlu}.

\begin{figure}[]
\centering
\hspace*{-0.0cm}\includegraphics[width=0.50\textwidth]{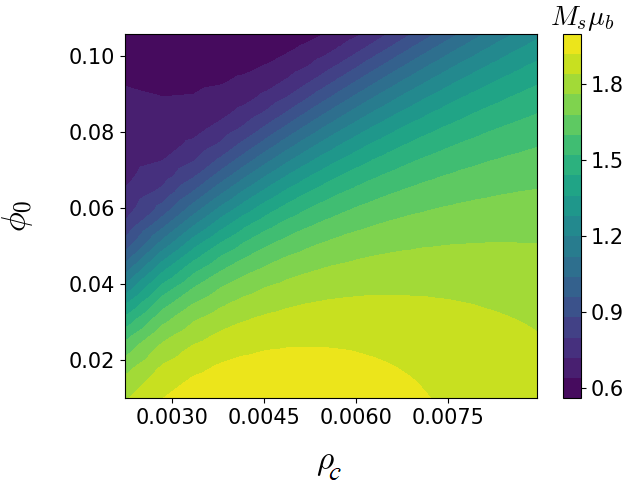}
\caption{Different total masses as functions of $\phi_0$ and $\rho_c$  for mixed static stars obtained with the polytropic values $K=100$, $\Gamma=2$ and self interacting coupling constant $\Lambda=100$.}
\label{mass_phi_rho_stat}
\end{figure}

A comprehensive plot of $M_s$ as a function of $(\phi_0,\rho_c)$ for $\Lambda=100$ is shown in~\cref{mass_phi_rho_stat}, aligned with this work's main results and previous research focused mainly on the static case. Despite the evident differences in the data setup, our static solutions, derived for high coupling constants, show positive comparability with the results presented in Fig.~1 from \cite{DiGiovanni:2021vlu}. In our study, \cref{mass_phi_rho_stat} exhibits the same behavior and ranges for both masses and the $\phi_0$ and $\rho_c$ parameters. The excellent agreement serves as an additional consistency check of our code.

Notably, the distinction between spinning and static bosonic matter is essential for comparison, as these configurations are inherently disconnected from one another due to differences in their matter distributions, morphology, and quantized angular momentum. Static bosonic stars are spherically symmetric objects, with the field central value corresponding to the maximum value of the field and playing a fundamental role alongside the internal field frequency parameter. In contrast, the central field value is zero for spinning configurations \cite{Yoshida:1997qf,Liebling:2012fv}, resulting in a toroidal mass distribution. Even the internal field frequency is very distinct between the two cases, being the eigenvalue obtained through the shooting method after fixing the central field in the static problem and the fundamental parameter of the model in the spinning case. However, we will try to compare both cases in the mixed setup briefly in the following section.

\subsection{Spinning solutions}\label{mainsol}

Spinning fermion-boson star solutions are complex and depend on numerous parameters, with various families of solutions exhibiting remarkable differences and features. We will focus on an illustrative family, leaving other cases for future work. Typically, spinning NS solutions are defined by fixed EOS parameters. For our polytropic models, the initial choices are the energy density at the origin $\bar{\rho}_0$ and the adiabatic index $\Gamma$. Additionally, an initial value for the rotational frequency should be provided. For our code, however, it is preferable to give an initial value for the oblateness parameter, $r_{ratio}$, instead. 
The static case corresponds to $r_{ratio}=1$ while our rotating stars are obtained for initial values $r_{ratio}=0.9$. This value, though, will vary slightly with the iterations of Ridder's method, as indicated in  \Cref{integration_method}. Different initial $r_{ratios}$ values will yield qualitatively similar but varying results. Pure NSs are easier to explore due to specific options in the original RNS code, allowing to fix particular solutions by setting final properties like mass, angular momentum, or axis ratios.

For rotating BS configurations, four main parameters must be determined: the boson mass, the self-interaction coupling constant, the harmonic index, and the internal field frequency $\bar{w}$.  Taking into account the re-scaling defined in \Cref{section3} and the dimensional values of  $G$, $c$ and $\hbar$, we obtain the following numerical values: $\bar{\mu}_b=0.01\rightarrow\mu_b\sim 1\cdot 10^{-16}\,\mathrm{MeV}$, $\bar{\eta}=0.01\rightarrow \eta=1$ and $\bar{\lambda}=1\rightarrow\lambda=100$. As the dynamical variable $\bar{w}$ imparts distinct characteristics to each mixed star. We present a family of solutions through systematic variations of initial parameters $\bar{w}$ and $\bar{\rho}_0$, while the other parameters are kept fixed and the final value of $r_{ratio}$ is obtained through the method for the given initial parameters $\bar{w}$ and $\bar{e}_c= \bar{\rho}_c + \bar{\rho}_i$, with this quantity being the total central mass-energy density. Our approach uses $\bar{e}_c$ instead of $\bar{\rho}_c$ due to the RNS output structure, allowing a comprehensive exploration of the parameter space, enhancing the understanding of  mixed stars properties within the theoretical framework.

It is relevant to mention that the solutions we discuss here mostly align with the most probable astrophysical scenario. In this scenario, a NS could begin to accrete varying amounts of bosonic matter, giving rise to systems with a well-defined inner NS 
with more or less bosons orbiting it until equilibrium configurations are reached. This  highlights the importance of this study within the broader context of astrophysical phenomena, in which NSs are well-known and significant.

Before delving into a comprehensive analysis of the entire suite of solutions, we propose an initial examination focused on the behavior of the fermionic matter, and the bosonic component across radial and angular dimensions. This examination will be conducted through the lens of four distinct models. This methodological approach enables a detailed inspection of the interplay between these critical aspects of the system, offering insights into how they vary spatially.

\begin{figure*}
    \centering
    \includegraphics[width=0.32\textwidth]{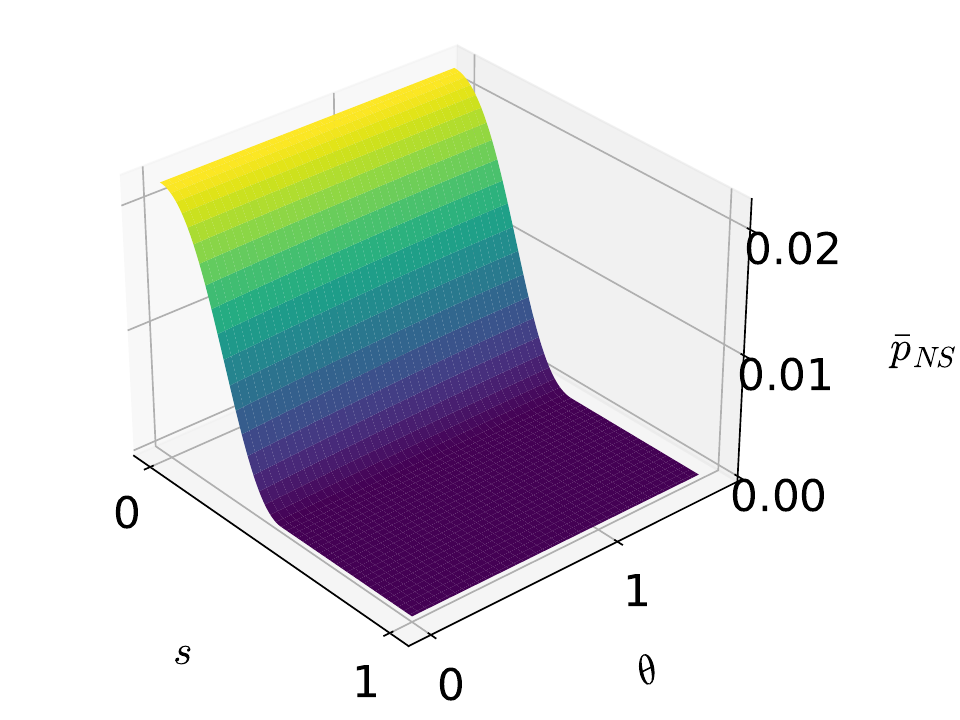}\hfill
    \includegraphics[width=0.32\textwidth]{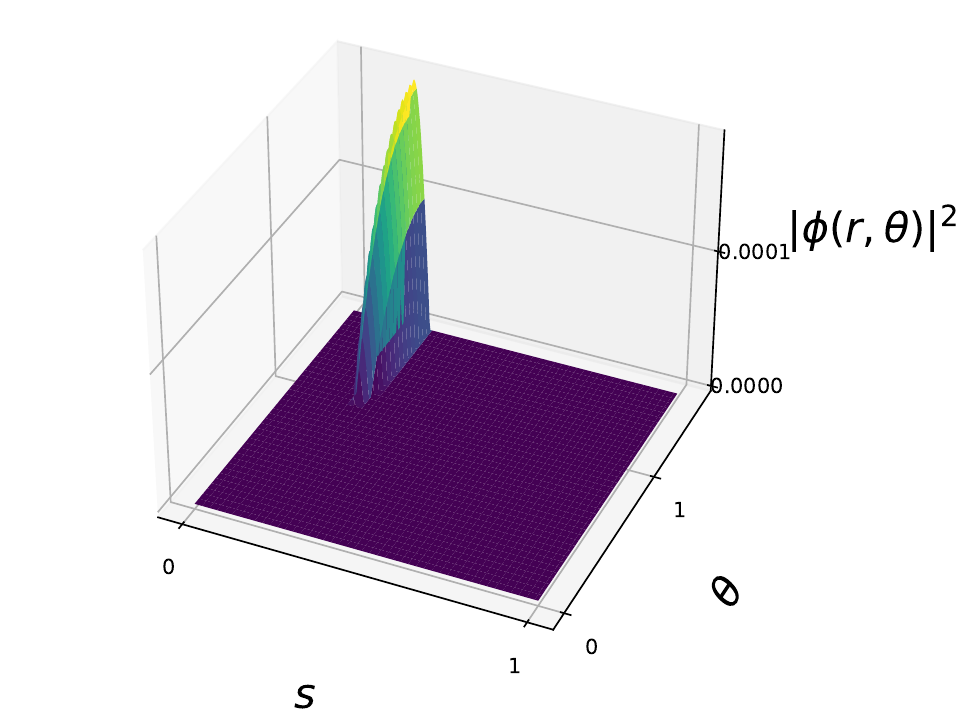}\hfill
    \includegraphics[width=0.32\textwidth]{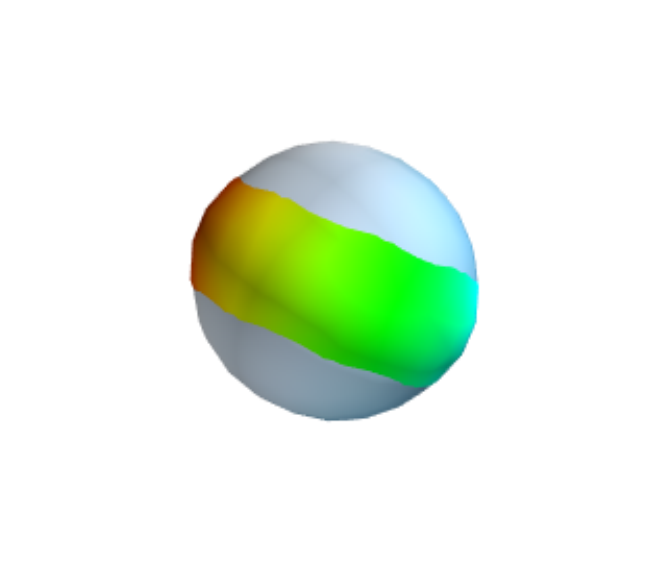}\label{fig:f1}
    
    \medskip

    \includegraphics[width=0.32\textwidth]{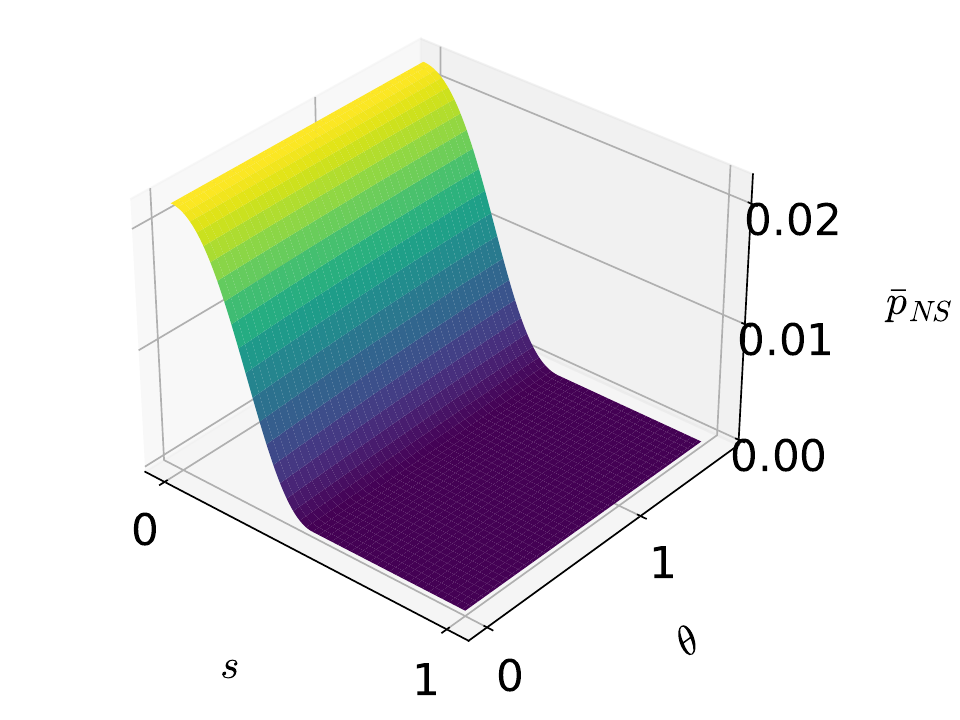}\hfill
    \includegraphics[width=0.32\textwidth]{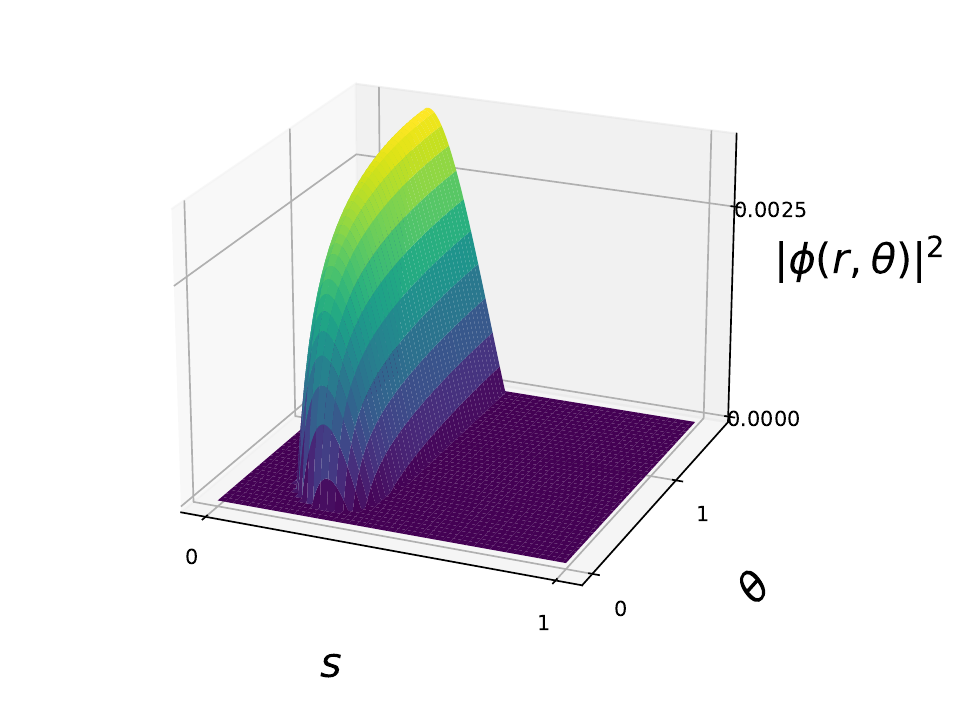}\hfill
    \includegraphics[width=0.32\textwidth]{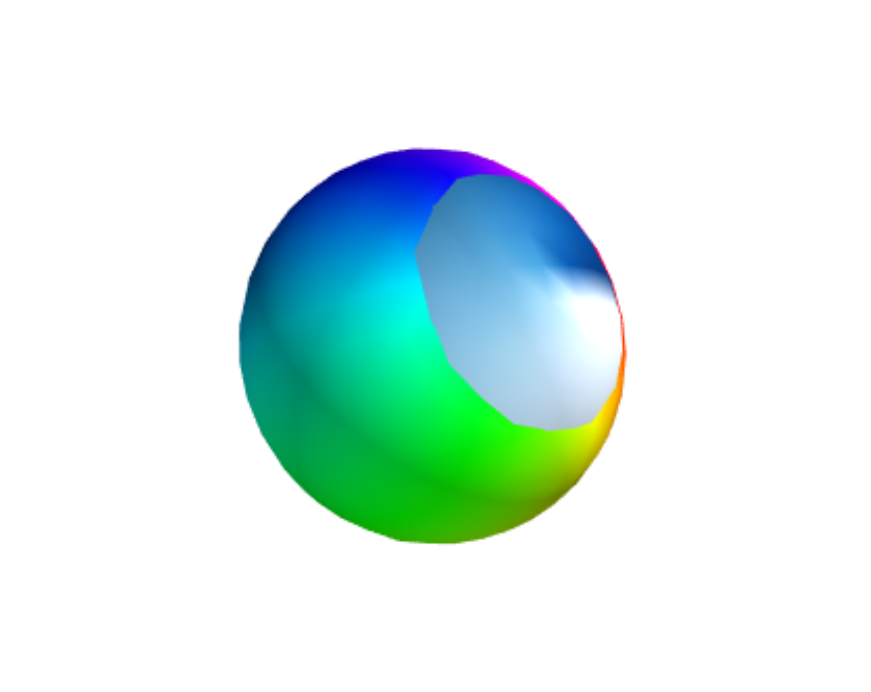}\label{fig:f2}
    
    \medskip

    \includegraphics[width=0.32\textwidth]{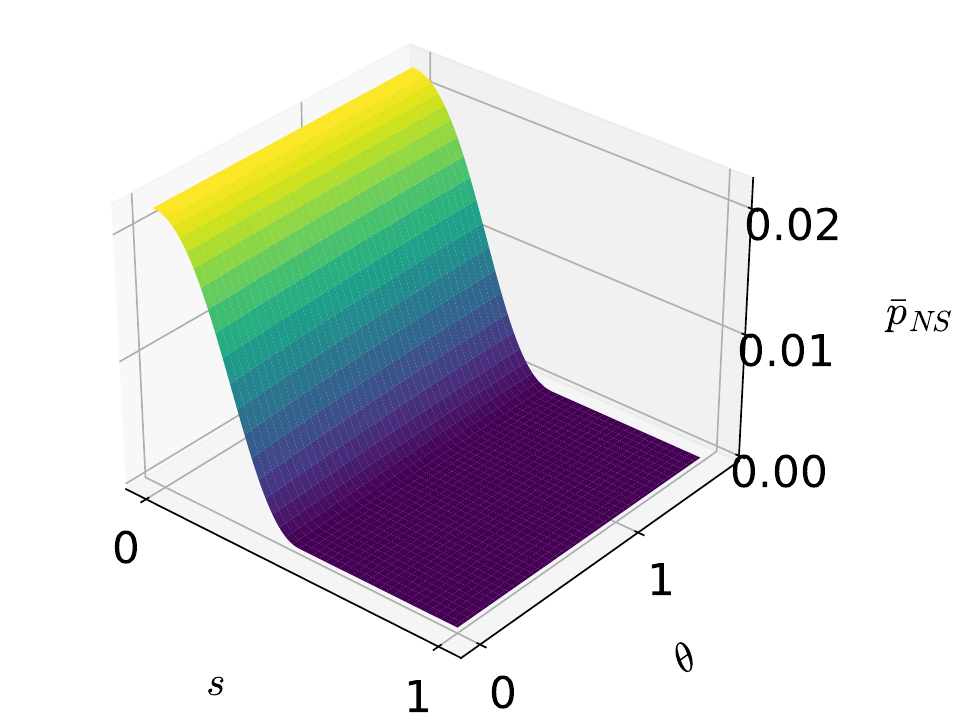}\hfill
    \includegraphics[width=0.32\textwidth]{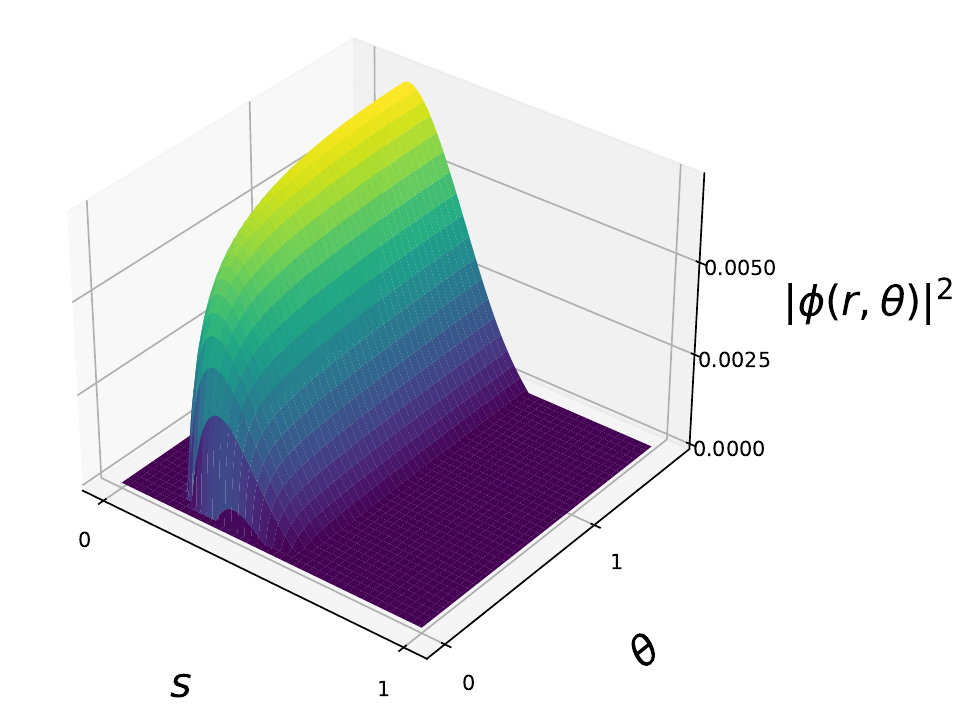}\hfill
    \includegraphics[width=0.32\textwidth]{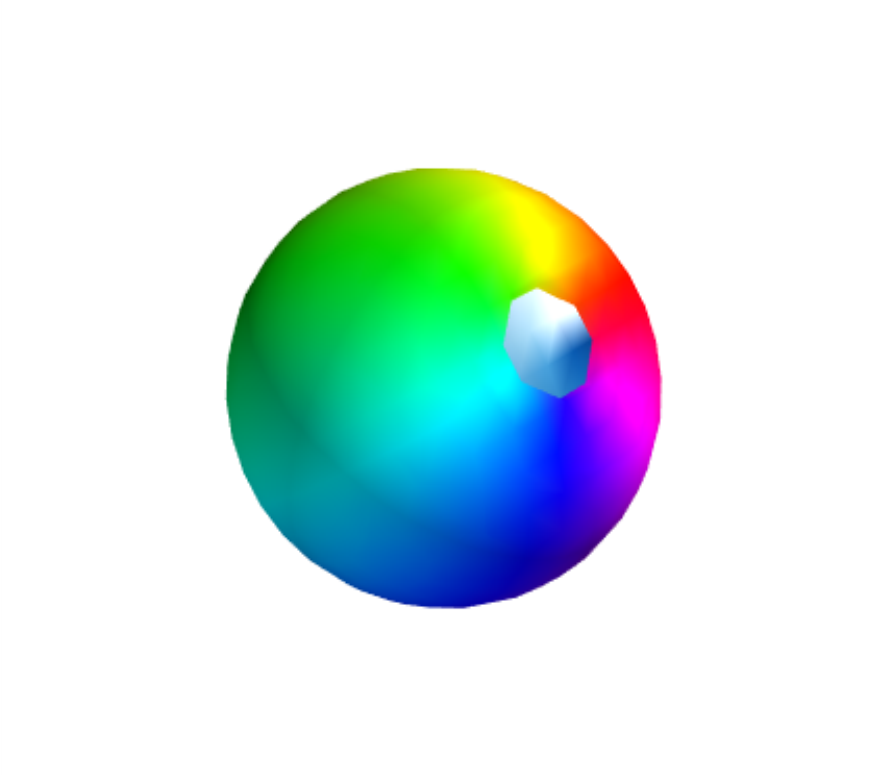}\label{fig:f3}
    
    \medskip

    \includegraphics[width=0.32\textwidth]{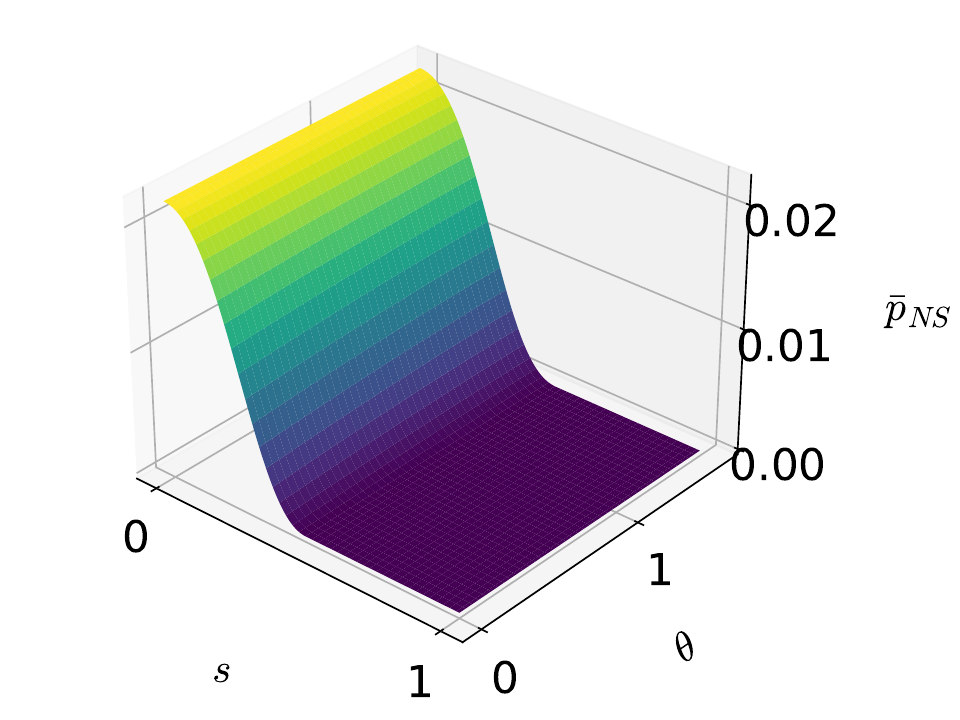}\hfill
    \includegraphics[width=0.32\textwidth]{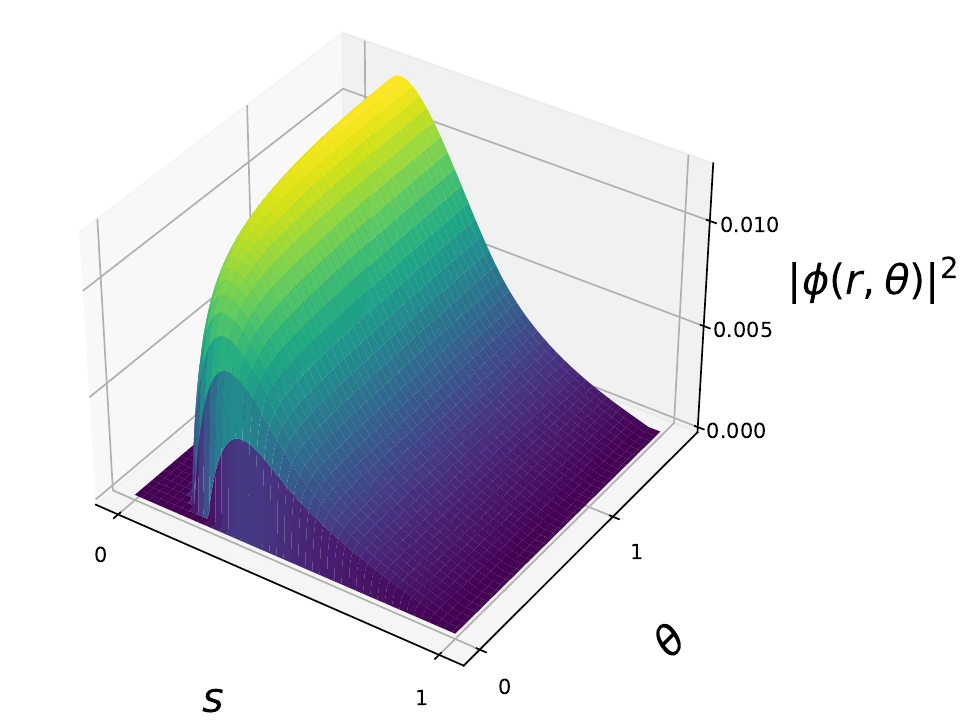}\hfill
    \includegraphics[width=0.32\textwidth]{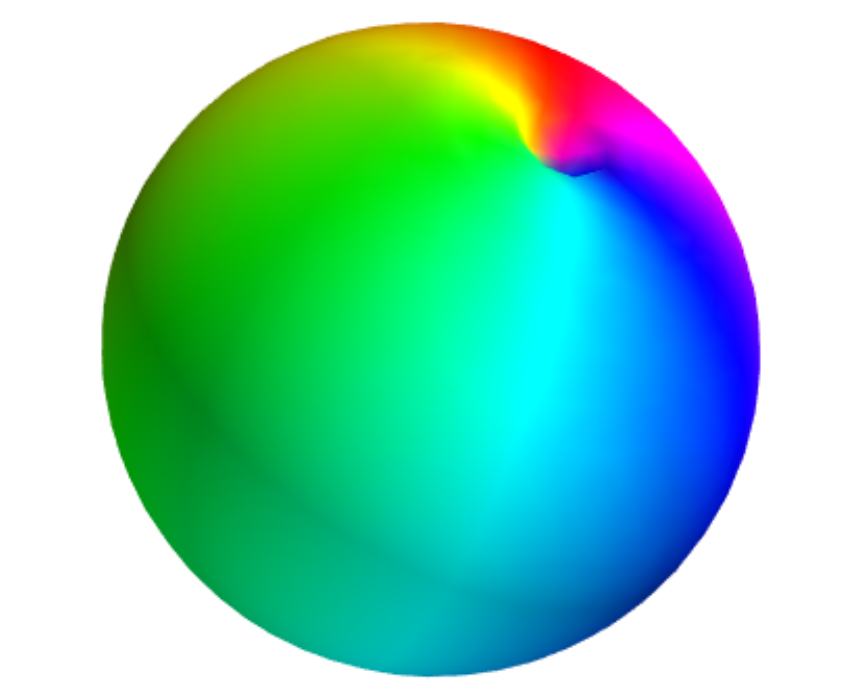}\label{fig:f4}
    
    \caption{Different mixed star solutions having the same central density $\bar{e}_c=0.17$ and $\Gamma=1.0$. The upper solution was obtained for  $\bar{w}=0.069$. The second row shows a $\bar{w}=0.078$ star. The third row shows a star with a frequency  $\bar{w}=0.088$. The last row is for the limiting case  $\bar{w}=0.0991$ star. It is important to notice that the radial coordinate is the compactified radius $s\in[0,1]$ and by columns, we plot $\bar{p}_{NS}(r,\theta),|\phi(r,\theta)|^2$ and the contour plots show the energy densities for both parts, being the grey/blue spheres, the fermionic part and the multi-color contour the field energy density $|\phi(r,\theta)|^2$. The color in the bosonic part represents the phase.}
    \label{fig4}
\end{figure*}

\begin{figure}[]
\includegraphics[clip,width=1.0\columnwidth]{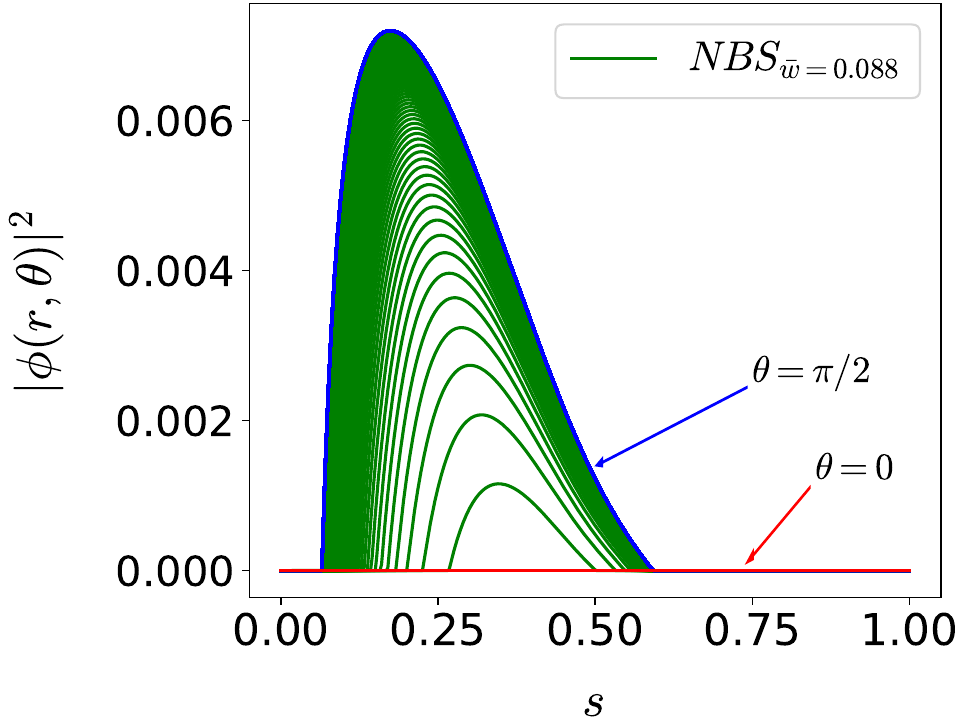}\\
\vspace{0.4cm}
\includegraphics[clip,width=1.0\columnwidth]{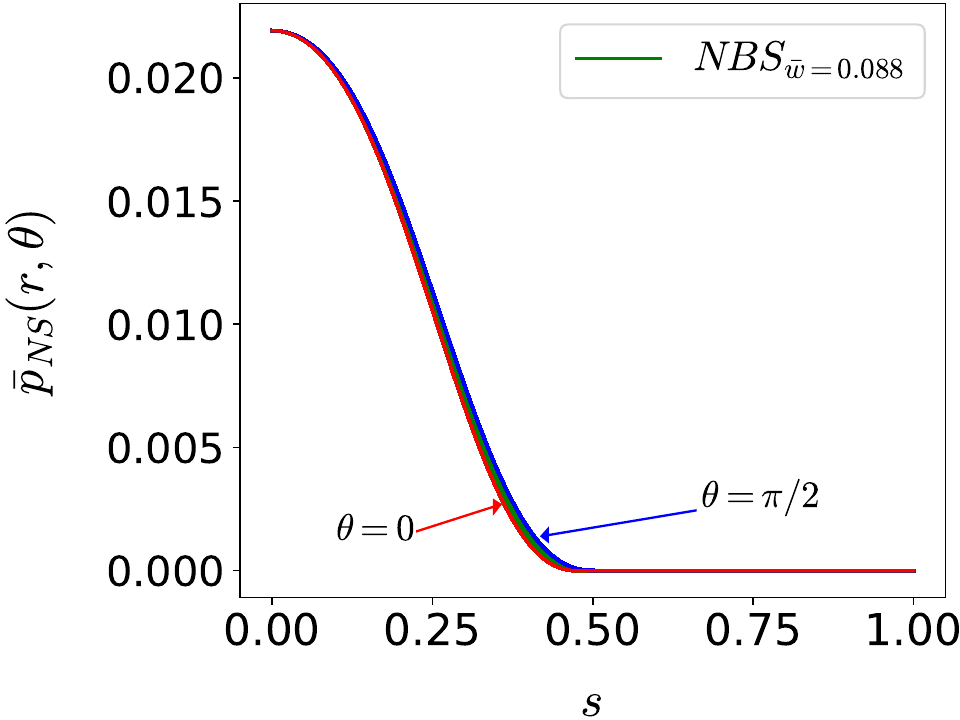}%
\caption{Radial profiles of the magnitude of the scalar field (top panel) and fermionic pressure $p_{NS}$ (bottom panel) for the spinning mixed star model with central energy density $\bar{\epsilon}_c=0.17$ and field frequency $\bar{\omega}=0.088$. The different lines correspond to different values of the angular coordinate $\theta$ with the red and blue lines corresponding to $\theta=0$ and $\theta=\pi/2$, respectively.  The red line in the upper plot shows that $|\phi(r,\theta)|=0$  at $\theta=0$, indicating a toroidal energy distribution of the scalar field.}
\label{profiles}
\end{figure}

From the analysis of \cref{fig4}, we deduce several notable properties of mixed stars within the framework of our study. Each row in the figure is dedicated to a distinct star, with subsequent columns providing detailed visualizations of the variables  $\bar{p}_{NS}(s,\theta)$ and $|\phi(s,\theta)|^2$ across the $s,\theta$ grid. The final column offers a representation of both fermionic and bosonic energy densities, rendered in Cartesian coordinates. This structured presentation facilitates a comprehensive comparison of the physical characteristics inherent to each star, delineating the spatial distribution of pressure fields and scalar field configurations. Moreover, the juxtaposition of fermionic and bosonic energy densities in the concluding column underscores the interplay between these fundamental components, highlighting their distribution and relative contributions within the composite structure of the mixed stars.


Even with the markedly different scenarios under consideration, the pressure within the neutron star, $\bar{p}_{NS}$, depicted in the first column of \cref{fig4}, does not exhibit a significant contrast between the different models. In \cref{profiles} we plot the radial profiles of the two components of the spinning NBS configuration with $\bar{\epsilon}_c=0.17$ and $\bar{\omega}=0.088$ (third row in \cref{fig4}) for several values of the coordinate $\theta$. While the neutron star profile does not depend significantly on $\theta$ since the star is close to being spherically symmetric, the scalar field profile is fundamentally different from the static cases shown in~ \cref{static}. Spinning scalar boson star have a toroidal morphology, which is clearly seen here.

\begin{figure}[]
\vspace*{0.1cm}
\centering
\hspace*{-0.0cm}\includegraphics[width=0.450\textwidth]{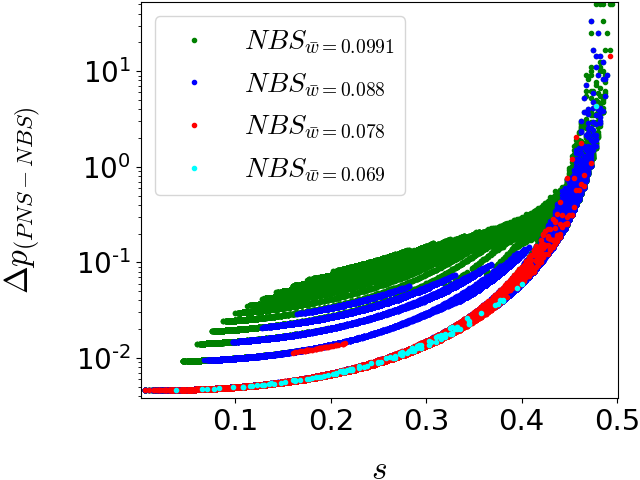}
\caption{Percentage difference between the pure NS and the four NBSs shown in \cref{fig4} plotted vs the computational radial coordinate.}
\label{pdif}
\end{figure}

Although the numerical values characterizing $\bar{p}_{NS}$ are distinctly different, it is difficult to visually distinguish the four cases shown in \cref{fig4}. In \cref{pdif}, however, we analyze the percentage difference between the purely NS case and the shown models. We make use of the following definition: 

\begin{equation}
    \Delta p_{(PNS-NBS)}=100\times\frac{|\bar{p}_{PNS}-\bar{p}_{NBS}|}{\bar{p}_{PNS}},
\end{equation}
where the sub-label $NBS$  refers to the mixed models and $PNS$ to the pure neutron star.

From \cref{pdif}, it becomes evident that models with higher field frequencies exhibit generally larger $\Delta p_{(PNS-NBS)}$ at any radial point. The plot also indicates that the greatest variation in neutron pressure occurs near the fermion surface, around $s \sim 0.5$. This demonstrates that the presence of the scalar field proportionally alters the fermionic matter: the greater the presence of the scalar field, the more significant the changes in the neutron pressure profile. However, it is in the representations of the scalar field and the energy density where the distinctions become more visually pronounced.

Upon scrutinizing the second and third columns
of \cref{fig4}, we observe a clear differentiation in the scalar field distributions and their corresponding energy density manifestations, analyzed on a row-by-row basis. The initial star demonstrates the most pronounced scalar field distributions, which, when interpreted in terms of energy density distribution, reveals a slender torus of bosonic matter that envelops and intersects the NS exterior.

In the subsequent scenario, the expansion of the field distribution is evident, indicating that the torus now encompasses a larger portion of the NS. This expansion is indicative of an increase in total mass. The third scenario illustrates a significant enhancement in the field's predominance to the extent that the NS is almost completely cloaked in bosonic matter. The final depicted scenario reveals a mixed star configuration where the NS is entirely encased in bosonic matter, with the torus expanding to such a degree that it nearly forms a spherical shape.

\begin{table}[h!]
\centering
\begin{tabular}{||c|c|c|c|c||}
\hline
\textbf{Models} & $\bar{\omega}$ & \textbf{Max $|\rho|$}  & \textbf{$M_{BS}/M_T$} &\textbf{$M_T$}$\left[M_{\odot}\right]$\\
\hline\hline
$NS$ &-& 0.86999 &0 &   1.5338 \\
\hline
$NBS_{\bar{w}=0.069}$ &0.069& 0.86999 &  3.9119$\cdot 10^{-7}$ & 1.5337\\
\hline
$NBS_{\bar{w}=0.078}$&0.078 & 0.86997 & 0.00021 & 1.5336\\
\hline
$NBS_{\bar{w}=0.088}$&0.088 & 0.87015 & 0.0018& 1.5346 \\
\hline
$NBS_{\bar{w}=0.0991}$&0.0991 &  0.88019 & 0.2580 & 2.0507\\
\hline
\end{tabular}
\caption{Maximum value of the metric function $\rho(s,\mu)$, mass ratio of the boson star, and total mass for the mixed models presented in~\cref{fig4}. The first row corresponds to the parameters of a pure NS with the same parameters $\bar{e_c}=0.17$, $r_{ratio}=0.9$. 
}
\label{tabledata}
\end{table}
In \Cref{tabledata} the maximum value for the metric function $\rho(s,\mu)$ is shown in the second column. While this quantity does not vary much as we increase the amount of scalar field, the total mass changes through our models, deviating from the NS. In particular, for the last case with $\bar{\omega}=0.0991$ the boson star mass represents a quarter of the total mass.


These observations stress the intricate interplay between the scalar field distributions and the resultant energy density surfaces, highlighting the transition from a slim torus to an almost spherical envelope of bosonic matter. This transition reflects the varying degrees of interaction between the fermionic and bosonic components.

Computing the total masses for the family of solutions as a function of the parameters  $M_{T}(\bar{w},\bar{e}_c)$ allows us to study the region 
where the scalar field couples more strongly, leading to more massive stars.

\begin{figure}[]
\includegraphics[clip,width=1.0\columnwidth]{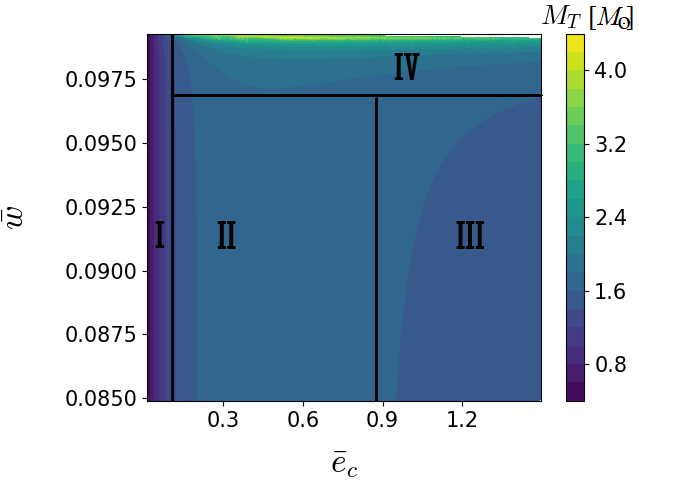}\\%
\includegraphics[clip,width=0.93\columnwidth]{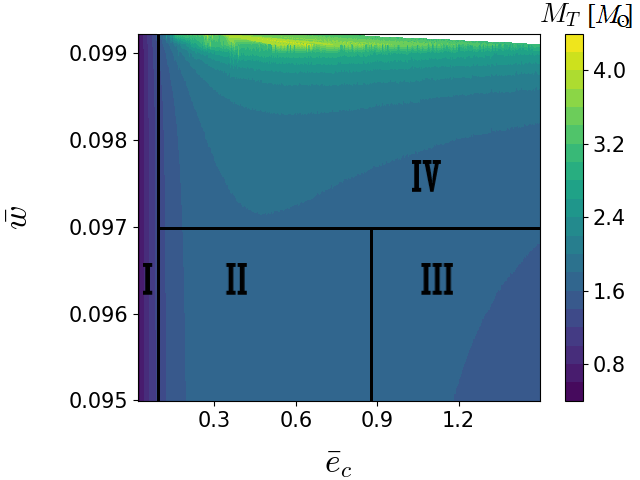}%
\caption{Contour plots depicting the total masses in units of solar mass as functions of the scalar field frequency and the central NS energy density are presented. The upper and lower plots display identical datasets, with the lower plot offering a magnified view of the region where the behavior of the bosonic component becomes more discernible.}
\label{masses}
\end{figure}

In \cref{masses} we represent solutions for the range of $\bar{w}$ between $0.085$ and $0.0993$. We explored, in fact, solutions for $\bar{w}\in [0.0685,0.0993]$, but for $\bar{w}<0.085$ the boson component contributes less than $2\%$ of the total mass, rendering them less relevant for our analysis.


We start the analysis by scrutinizing the upper plot in \cref{masses}, wherein the contour can be explained by splitting it into four principal regions. The first region is characterized by an exceedingly low NS energy density, where $\bar{e}_c \sim (0,0.1)$. Solutions within this parameter space represent the existence threshold for NS, manifesting very low masses and failing to meet the requisite density and compactness criteria essential for sustaining a scalar field. This segment of the plot can be elucidated by drawing a parallelism with  a mass-radius plot for NSs, specifically focusing on the segment where the mass is below $0.5 M_{\odot}$ and the radii exceed 13 km. Additionally, an inspection of this region with respect to $\bar{w}$ reveals negligible variations despite alterations in the parameter, aligning with the inference that the solutions are derived under a scenario dominated by fermionic matter, even if the fermioninc component is itself small. This suggests that a certain mass threshold for NSs is imperative to facilitate a significant coupling with the scalar field.

The second region is distinctly demarcated by the contours, encompassing the solution space bounded by $\bar{w}\in(0.068,0.097)$ —excluding certain solutions at the lower limit of $\bar{w}$ to omit redundant information —and $\bar{e}_{c}\in(0.1,0.85)$. This domain exhibits solutions with neutron star masses in the range of $1.8-2.3~M{\odot}$. Notably, the solutions within this region display low sensitivity to variations in frequency and energy density, indicating that a broad parameter space converges to similar types of solutions. This implies a certain robustness or invariance in the NS characteristics within this specific range of parameters, reflecting an area of uniformity in the solution space.

The third region is situated in the right portion of the upper plot in \cref{masses}. It is characterized by a decrease in the total mass beyond $\bar{e}_c \sim 0.9$. This trend aligns with observations from prior studies on polytropic NSs and mixed fermion-boson stars (see for instance~\cite{DiGiovanni:2020frc}) and was anticipated in our research. The influence of the field frequency on this behavior is evident, as the contours exhibit rounded borders, indicating a variation in mass with frequency. Analysis of this plot reveals that higher frequencies correlate with more massive solutions, which can be attributed to the increased coupling of the scalar field within the system, thereby leading to an increase of the total mass of mixed stars in this region.

Finally, the fourth region is the \textit{high frequency} parameter space ranging almost all the data based on the $\bar{e}_c$. We define it for solutions with $\bar{w}\geq 0.097$ and $\bar{e}_c >0.1$. This region covers a wide range of data predicated on $\bar{e}_c$ and is prominently illustrated in the lower panel of \cref{masses}. Several critical insights can be derived from this solution domain.

\begin{figure}[]
\includegraphics[clip,width=1.0\columnwidth]{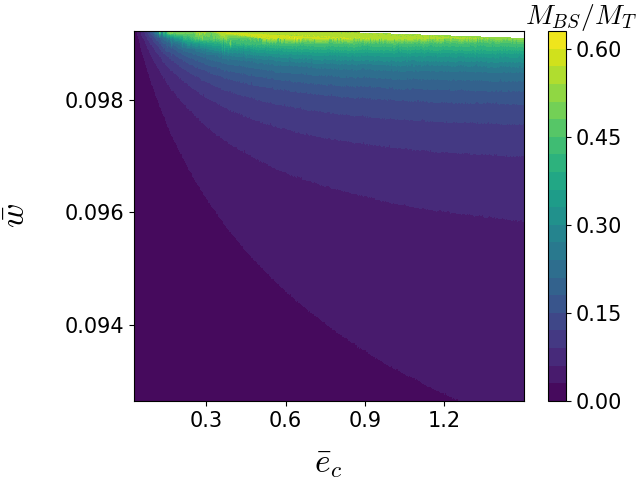}\\
  \includegraphics[clip,width=1.0\columnwidth]{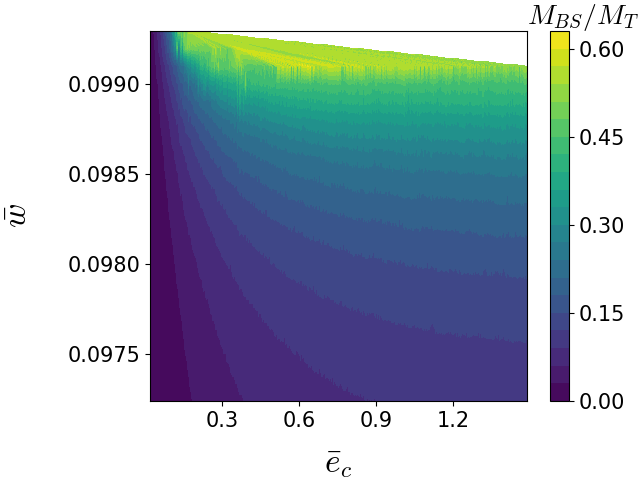}%
\caption{We show the relation between the bosonic and total mass for our mixed stars data set. The upper and lower plots have the same data, but zoom the region of interest in the second one.}
\label{massescomp}
\end{figure}

A pivotal observation is the attainment of maximum masses exceeding $4.5M_{\odot}$ within a small subregion, indicating the presence of exceptionally massive mixed star solutions. However, it is noted that these maximal mass values are not located at the extreme boundaries of the parameter space, as one could naively expect. Instead, there is a discernible decrease in mass beyond a specific neutron star energy density and scalar field frequency threshold. This behavior is visually represented by a narrow yellow band, flanked by green and blue regions, signifying that an increase in $\bar{w}$ and $\bar{e}_c$ beyond specific limits reduces the total mass. This phenomenon suggests a complex relation between the two significant parameters within this high-frequency region.

Another surprising behavior is found concerning the field frequency boundary. 
When increasing  $\bar{e}_c$, the space of $\bar{w}$ where the code finds equilibrium solutions decreases linearly. Further, in this region an effect already described for the third region is stronger. We see how the contours describe a sort of central higher mass region that clearly decreases after going beyond the thresholds. 
This pattern suggests the formation of a peak in the mixed star mass that decreases on either side of these thresholds, illustrating a nuanced interplay between central energy density and field frequency in determining the total mass.



Additionally, 
higher values of $\bar{w}$ lead to a convergence of contours, 
making this region richer in terms of the variety of solutions.
This behavior emphasizes the significant influence of the bosonic component on the equilibrium and structural characteristics of mixed stars within the high-frequency domain. 

A plausible explanation for this observed behavior could be as follows: our fermionic models involve rapid rotation, but generally at velocities that are lower than those seen in purely spinning bosonic stars. An increase in 
$\bar{w}$, which denotes a lower rotational frequency for bosonic stars, suggests a potential synchronization in rotational frequency between the fermionic and bosonic components. This synchronization could enable the mixed star to achieve higher total masses and better integration of both matter types. Specifically, more effective coupling occurs when the fermionic and bosonic components spin at comparable speeds, which is more likely to occur when the bosonic component has a lower velocity and, consequently, a higher 
$\bar{w}$. A more in-depth investigation of this phenomenon will be addressed in future research.

With the above considerations, we can compare the spinning with the static configurations shown in \cref{mass_phi_rho_stat}, noting the inherent differences between static and spinning scalar boson stars' energy density distributions.  The spinning case shows a richer structure, exhibiting distinct behaviors across sub-regions $I, II, III, IV$, and generally showing an increase in mass with higher internal frequencies. Conversely, the central scalar field parameter $\phi_0$, which plays the same role in the static case as the frequency, behaves differently. Notably, the total masses differ significantly between the two cases, with the static system achieving approximately $M_T\sim2 M_{\odot}$, while the spinning systems reach around $M_T\sim4.5 M_{\odot}$. However, there is a similar trend in both cases: beyond a certain value of the central neutron energy density, the total mass decreases, as the object cannot accommodate any more matter.

The subsequent \cref{massescomp} shows the ratio between the mass of the bosonic part and the total mass obtained by summing the fermionic and bosonic contributions. This figure helps to clarify and expands on some of the above conclusions. The upper panel shows a broader range of solutions where we can identify solutions with less than a $10\%$ of bosonic content, shown in dark blue, and solutions where the BS part represents more than $60\%$ of the total mass in light green and yellow.
It becomes clear from this plot that the \textit{medium-high energy density}-\textit{high-frequency} region is the part of the parameter space where the bosonic contribution becomes important.

We can go even further in the analysis if we zoom into the above-mentioned region of interest.
The lower panel in \cref{massescomp} shows this region in more detail. The first assessment we can make is related to the high variability of the space of solutions, meaning that the solutions in this area are strongly affected by any variation in $\bar{w}$. We also checked that the spectra of solutions decrease for the frequency when $\bar{e}_c$ increases.
Additional comments can be made about the maximal bosonic mass found in our data set. There is a tiny region for non-extremal high frequencies and medium NS energy densities, where the bosonic component reaches values above $60\%$ of the total mass. Moving beyond these values of the parameters, the bosonic contribution to the total mass decreases significantly.

It is important to acknowledge that the study of the higher spectrum of solutions is constrained by potential numerical challenges. Our computational model is designed to solve problems within specific boundary conditions. However, an unregulated increase in the number of bosons could lead to a scenario where asymptotic flatness is violated. In such cases, the code struggles to converge to a meaningful equilibrium solution as the  fundamental assumptions of the model are violated. This is evident in the upper right corners of \Cref{masses} and \Cref{massescomp}, where a blank region indicates that the code does not find any solution.

This limitation underscores the intricate balance required in numerical simulations, especially when dealing with systems where the interplay of various components —such as the neutron star and boson star elements— can lead to complex and, at times, unpredictable outcomes. The inability to reach a broader spectrum of solutions highlights the system's sensitivity to boundary conditions and the need for a careful control of parameters to ensure the reliability and relevance of the simulated solutions.

\subsection{Presence of ergoregions}\label{ergoregions}

The ergoregion instability appears in any system with
ergoregions and no horizons. For some models of rapidly rotating NS, it was shown that ergoregions can arise when dealing with certain EOS \cite{Tsokaros:2019mlz} where typical
instability time scales are shown to be larger than the
Hubble time.  In these cases, the ergoregion instability is too
weak to affect the star's evolution. This conclusion changes drastically for ultracompact stars.
Depending on the compactness, instability time scales ranging from seconds to even weeks were found \cite{Cardoso:2007az}. Indeed, a rigorous study was performed for various types of exotic compact objects \cite{Maggio:2018ivz}. Furthermore,   BSs have been proven to support ergoregions depending on the self-interaction governing the field, since this parameter is intrinsically linked to the level of compactness achievable by models featuring self-interactions in the potential \cite{Vaglio:2022flq}. Those fast-spinning, massive BSs are directly related to our mixed stars.

\begin{figure}[]
\centering
\hspace*{-0.0cm}\includegraphics[width=0.50\textwidth]{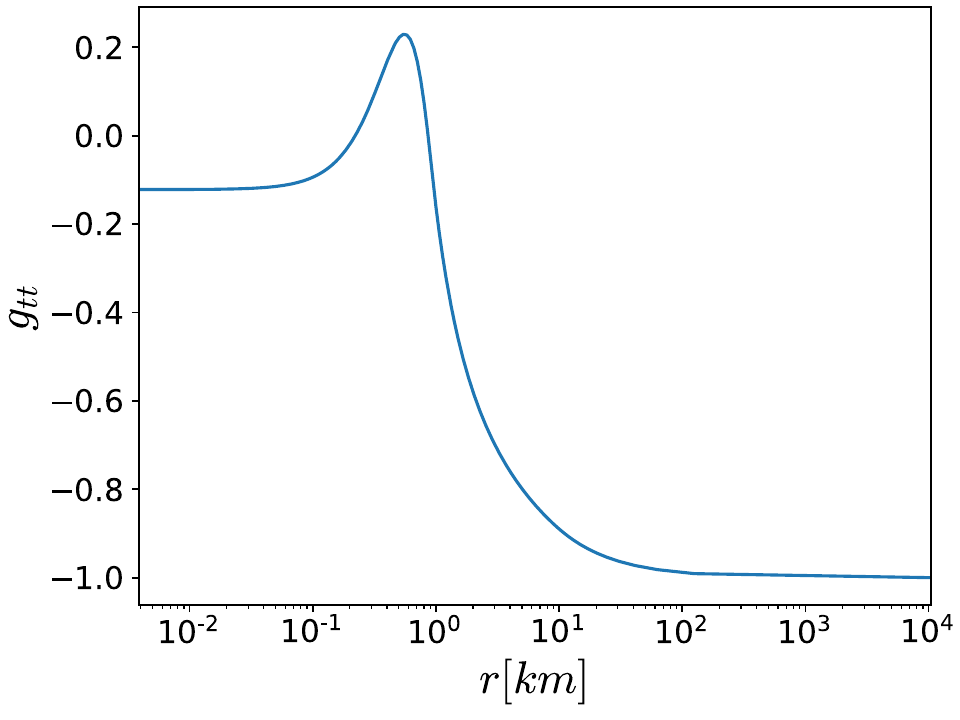}
\caption{For the 
EOS presented in \cite{Adam:2020yfv} and some very specific parameters, we have found ergoregions with toroidal topology in our mixed star solutions. These manifest as a change in the $g_{tt}$ sign. }
\label{ergo}
\end{figure}

On the other hand, the mixed stars we have obtained with the polytropic EOS discussed in this work, with $\Gamma=1.0 $, and for any value of $\bar{w}$ did not present any ergoregion. However, considering different polytropic indexes resulted in drastic changes in the compactness of our solutions, and some of them showed this kind of region where no particle can remain at fixed $r,\theta$ and $\psi$.
To further underline this point, and going beyond the polytropic EOS, 
we present in \cref{ergo} a solution obtained for a rather realistic, tabulated EOS \cite{Adam:2020yfv} where we appreciate two changes in the sign of  $g_{tt}$, which translates into the existence of an ergoregion, demonstrating that mixed stars can also support ergoregions. Firstly, we show their existence for some of our results, making a clear connection with \cite{Tsokaros:2019mlz,Vaglio:2022flq}. Secondly, this result anticipates a future research line since our solver is capable of using tabulated EOS. In future work, therefore, we will be able to analyze in detail and discuss more realistic models from the point of view of the NSs. 

\section{Fermion-Boson stars in gravitational-wave astronomy}
\label{section6}

While alternative scenarios have been proposed \cite{Bustillo:2020syj,bustillo2023searching}, the vast majority of current GW detections of compact binaries are perfectly described by that of mergers of black holes and/or NS \cite{GWTC1_PRX,abbott2021gwtc2,GWTC2.1,abbott2021gwtc3}. There exist, however, observations that challenge the assumed characteristic masses of these objects. For instance, the event GW190814 \cite{GW190814} involves a compact object of around $2.6M\odot$, falling in the so-called ``lower-mass black-hole gap'', where the formation of black holes from stellar-collapse is prevented. While scenarios involving primordial black holes \cite{Vattis2020,GW190814_Bellido} and dark matter \cite{lee2021could,Das2021} have been proposed to explain such events, fermion-boson stars can also populate this mass range, offering an alternative scenario~\cite{DiGiovanni:2021ejn}. While we are far from performing actual numerical simulations of mergers of these objects, we anticipate two potential avenues to confirm or discard such an option. First, GWs emitted during the end stages of the inspiral phase will contain information about the characteristic tidal deformability of these objects. Second, the bosonic part of the star will affect both the nature of the final object, most likely making it collapse into a black hole due to the increased mass of the binary in comparison to the case of pure NS binaries, and any potential post-merger emission. In both cases, third-generation detectors such as Einstein Telescope \cite{ET1,ET2}, Cosmic Explorer \cite{CE,CE2} or NEMO \cite{Ackley2020} will be needed to access the corresponding information.

The expanded array of options provided by our mixed solutions offers essential insights that could address the inconsistencies observed in recent multi-messenger observations and nuclear physics experiments concerning NS masses and radii. Notably, our developed solutions exhibit significant coherence with multi-messenger data, e.g., by populating the lower-mass black hole gap with our hypothetical compact fermion-boson stars, in the line with the discussion in \cite{DiGiovanni:2021ejn}. This coherence extends beyond the previously mentioned GW to include X-ray pulsars PSR J0030+0451 \cite{Miller:2019cac} and PSR J0740+6620 \cite{Miller:2021qha}.

We note that the above statements, including the viability of isolated rotating mixed stars and of compact mergers involving fermion-boson stars, crucially depend on a few assumptions that we outline next.
Despite the positive outcomes achieved and the potential relevance to the mass-gap context, it is imperative to exercise a degree of caution due to the specific and limiting assumptions that were applied to derive our solutions. First of all, for the stars composed of fermionic matter we mainly employed polytropic EOS in the present work rather than other tabulated, more realistic EOS. This implies that our models should be considered approximations from the outset. Nonetheless, as we have indicated in the previous section, and  in spite of not presenting them in this work, our solver can  readily be generalized to accommodate tabulated EOS. 
Undoubtedly, the most significant limitation arises from employing the Ryan-Pani-Vaglio approximation for the scalar field \cite{Vaglio:2022flq,Ryan:1996nk}. 





This approach restricts the range of values of the coupling constants or the adoption of different potentials, making the total masses obtained through our algorithm bounded by the method. Although we cannot guarantee that the use of other potentials or coupling constants within another algorithmic framework will not quantitatively alter the solutions described here, it is likely that this would allow exploring further total mass ranges and comparing mixed stars with different astrophysical observations outside the mass-gap regime. Besides,
as explained in \Cref{mixedrotstars}, numerical studies on the stability of bosonic stars~\cite{DiGiovanni:2020ror,Dmitriev:2021utv,Siemonsen:2020hcg,DiGiovanni:2021vlu,sanchis2021multifield} support our hypothesis on the inherent stability of these objects. This is inferred from the fact that both the fermionic and bosonic components, when examined independently, have demonstrated sufficient dynamical stability. However, this needs to be proven in numerical time evolutions of such objects, which will be explored in future work. Therefore, this approximation is justifiable and relevant, as it allows us to explore new potential candidates in mass-gap events.

In conclusion and by way of anticipation, we think that, despite its approximate nature, our model is likely not far removed from an exact representation. Solving the complete system with realistic EOS and integrating the scalar field without any approximation will undoubtedly alter to a certain degree the total mass and the relative proportions between fermionic and bosonic masses. However, we posit that the nature of the solutions will closely resemble our findings, both in terms of the object's geometry or topology and the total magnitude or range of the masses.

\section{Conclusions}
\label{section7}

We have presented a new kind of hypothetical compact object: a fully spinning mixed fermion-boson star with a specific bosonic self-interaction, valid within the large self-coupling regime. The Einstein equations for the axisymmetric system, sourced by a mixed stress-energy tensor \cref{fulleinstein}, have been solved by a $\mathcal{C}$-code that takes the RNS \cite{saphiro} as a base. Due to the large number of parameters defining this kind of system, the available zoo of solutions is large. For this reason, we present only one complete family of configurations, and study one of their main properties, which is the total mass, as it is the first time this kind of object is treated in depth. 

We have delineated the theoretical framework incrementally before integrating it into a coherent system to elucidate our conceptualization of the hypothetical astrophysical object. Subsequently, we systematically detail the implementation of the intricate algorithm employed for solving this integrated system.

Following the derivation of accurate expressions intended to quantify observable magnitudes, we conducted simulations of several representative spinning cases. These simulations, grounded in carefully selected initial guesses, facilitated the exploration of the most plausible astrophysical scenarios. Specifically, these scenarios involve a NS that variably accretes bosons, thereby illustrating the nature of such astrophysical objects. Not only were the metric potentials and scalar field behaviors elucidated for a selection of distinct and representative cases, but volumetric representations of energy densities for both types of matter were also constructed and analyzed. These visualizations provide insights into the comprehensive matter distributions of these astrophysical entities within spatial dimensions. Furthermore, the analysis allows for the observation of a conventional, flattened spherical spinning NS encased within a toroidal distribution of the scalar field, highlighting the complex interplay between the NS and surrounding scalar fields.


By fixing specific parameters such as the oblateness of the inner NS, the harmonic index, polytropic variables, and the boson particle mass, we derived a comprehensive suite of solutions by adjusting the internal bosonic scalar field frequency $\bar{\omega}$ and the central energy density of the NS $\bar{e}_{c}$. 
To construct these solutions, we begin by taking an initial guess that assumes a static neutron star as explained in \cref{section3B}. The resulting configurations are richer in fermionic content at low frequencies and lower central energy densities $\bar{e}_c$, while bosonic matter becomes more significant in certain constrained regions. Generally, this emphasis on the fermionic component is motivated by the fact that it is considered the most likely astrophysical scenario.


Our findings suggest that the novel astrophysical entities introduced in this study are capable of achieving greater masses in some scenarios, particularly when the frequency 
$w$ closely approaches but remains smaller than the value of 
$\mu_b$, and when the central energy density increases within specific bounds, rather than arbitrarily. Consequently, it becomes feasible to form objects with masses surpassing $3M_{\odot}$, potentially extending up to $4M_{\odot}$.


Our analysis also compares the contribution of bosonic matter to the total mass, highlighting the impact of the amount of scalar bosonic matter present in the mixed star. Even though fermionic matter usually dominates in areas with low field frequencies and central NS energy densities, we notice that the proportions of bosonic and fermionic matter balance out when the field frequency is high and when $\bar{e}_c$ increases. Interestingly, we find some models for which more than $50\%$ of the mass comes from the bosonic component. Our investigation further extended to the presence of ergoregions, yielding positive outcomes. These findings were realized after the application of an advanced, albeit not comprehensively examined, methodology involving the use of realistic EOS during the integration process of the mixed system.

One of the most significant contributions of our results could lie in their potential to elucidate certain astrophysical phenomena currently interpreted as mass-gap black holes, specially among GW observations. The compact objects detailed in this manuscript, though not entirely realistic owing to the adoption of polytropic models and large self-interaction constants for NS and BS components, respectively, illuminate a novel avenue for extending these models to more encompassing and plausible scenarios. Our analysis, bolstered by additional verifications not detailed herein, suggests that incorporating realistic EOS could diversify the spectrum of solutions. This includes the possibility of achieving higher masses in certain instances, altering the scalar field configurations, and fostering varied fermion-boson equilibria, wherein the NS component primarily functions as a core. However, given the ongoing nature of our investigation, we reserve more bold conclusions for future work.

Our ongoing research endeavors are directed towards intriguing expansions of our previous work. Notably, the incorporation of realistic EOS has been identified as a viable direction, though it remains under-explored in a systematic context. An immediate enhancement under consideration involves transcending the approximation of large self-interaction coupling constants to address the dynamics of a fully realized scalar field, integrated via a generalized Stress-Energy tensor. Furthermore, the exploration of varied self-interaction potentials for the BS component presents a fertile ground for additional inquiry. The extension of our methods to encompass spinning fermion-Proca stars is another avenue of interest. There are also promising insights into the study of stability and evolutionary trajectories of these novel mixed compact objects. Moreover, the investigation of Universal Relations, the multipolar structure, and the determination of deformability parameters for these entities are anticipated to yield significant implications. Such advancements not only augment our understanding of compact astrophysical objects but also enhance the theoretical framework necessary for interpreting their observational signatures.

\begin{acknowledgements}
 JCM thanks Jos\'e A. Font, F.Di Giovanni, E. Radu, C. Herdeiro, A. Wereszczynski, M. Huidobro, and A.G. Martín-Caro for some crucial discussions. 
Further, the authors acknowledge financial support from the Ministry of Education, Culture, and Sports, Spain (Grant No. PID2020-119632GB-I00), the Xunta de Galicia (Grant No. INCITE09.296.035PR and Centro singular de investigación de Galicia accreditation 2019-2022), the Spanish Consolider-Ingenio 2010 Programme CPAN (CSD2007-00042), and the European Union ERDF.
 JCM thanks the Xunta de Galicia (Consellería de Cultura, Educación y Universidad) for the funding of their predoctoral activity through \emph{Programa de ayudas a la etapa predoctoral} 2021. JCM thanks the IGNITE program of IGFAE for financial support. JCB is funded by a fellowship from ``la Caixa'' Foundation (ID100010474) and from the European Union's Horizon2020 research and innovation programme under the Marie Skodowska-Curie grant agreement No 847648. The fellowship code is LCF/BQ/PI20/11760016. JCB is also supported by the research grant PID2020-118635GB-I00 from the Spain-Ministerio de Ciencia e Innovaci\'{o}n. NSG acknowledges support from the Spanish Ministry of Science and Innovation via the Ram\'on y Cajal programme (grant RYC2022-037424-I), funded by MCIN/AEI/10.13039/501100011033 and by ``ESF Investing in your future”. NSG is further supported by the Spanish Agencia Estatal de Investigaci\'on (Grant PID2021-125485NB-C21) funded by MCIN/AEI/10.13039/501100011033 and ERDF A way of making Europe, 
and by the European Horizon Europe staff exchange (SE) programme HORIZON-MSCA2021-SE-01 Grant No. NewFunFiCO-101086251.
\end{acknowledgements}

\bibliography{biblio}


\begin{appendix}

\begin{widetext}

\section{Sources}
\label{appendixA}


We introduce a series of equations that, being of main interest, we do not add to the text for clarity of reading.
First of all, we show the sources for the spinning fermion-boson case:

\begin{equation}
\begin{split}
    &S_{\rho}^T(r,\mu)= e^{\gamma/2}\left[8\pi e^{2\alpha}\left\{\left(\rho_0+\rho_i+p_{NS}\right)\frac{1+v^2}{1-v^2}+\left(\rho_{BS}+p_{BS}\right)\frac{1+\Bar{v}^2}{1-\Bar{v}^2}\right\}+r^2(1-\mu^2)e^{-2\rho}\left(\omega_{,r}^2+\frac{(1-\mu^2)}{r^2}\omega^2_{,\mu}\right)+\right.\\
    &\left.\frac{1}{r}\gamma_{r}-\mu\gamma_{,\mu} +\frac{\rho}{2}\left\{16 \pi e^{2\alpha}\left(p_{NS}+p_{BS}\right)-\gamma_{,r}\left(\frac{1}{2}\gamma_{,r}+\frac{1}{r}\right)-\frac{1}{r^2}\gamma_{,\mu}\left(\frac{1-\mu^2}{2}\gamma_{,\mu}-\mu\right) \right\}\right],
\label{Sr}
\end{split}
\end{equation}

\vspace{1cm}

\begin{equation}
    S_{\gamma}^T(r,\mu)=
    e^{\gamma/2}\left[16\pi e^{2\alpha}\left(p_{NS}+p_{BS}\right)+\frac{\gamma}{2}\left\{16\pi e^{2\alpha}\left(p_{NS}+p_{BS}\right)-\frac{1}{2}\gamma_{,r}^2-\frac{1-\mu^2}{2r^2}\gamma_{,\mu}^2\right\}   \right],
\label{Sg}
\end{equation}

\vspace{1cm}

\begin{equation}
\begin{split}
    &S_{\omega}^T(r,\mu)=e^{(\gamma-2\rho)/2}\left[-16\pi e^{2\alpha}\left[\frac{(\Omega-\omega)(\rho_0+\rho_i+p_{NS})}{1-v^2}+\frac{\Bar{v}\left(\rho_{BS}+p_{BS}\right)}{(1-\Bar{v}^2)r\sin\theta}\right]+\right.\\
    &\left.\omega\left\{ -8\pi e^{2\alpha}\left[\frac{(1+v^2)(\rho_0+\rho_i)+2v^2 p_{NS}}{1-v^2}+\frac{(1+\Bar{v}^2)\rho_{BS}+2\Bar{v}^2 p_{BS}}{1-\Bar{v}^2}\right] -\frac{1}{r}\left(2\rho_{,r}+\frac{1}{2}\gamma_{,r}\right)+\frac{\mu}{r^2}\left(2\rho_{,\mu}+\frac{1}{2}\gamma_{,\mu}\right)+ \right.\right.\\ 
    &\left.\left. \frac{1}{4}(4\rho_{,r}^2-\gamma_{,r}^2)+\frac{1-\mu^2}{4r^2}(4\rho_{,\mu}^2-\gamma_{,\mu}^2)-r^2(1-\mu^2)e^{-2\rho}\left[\omega_{,r}^2+\frac{(1-\mu^2)}{r^2}\omega_{,\mu}^2\right]    \right\}\right],
\label{Sw}
\end{split}
\end{equation}

\begin{equation}
\begin{split}
& \alpha^T_{, \mu}=-\frac{1}{2}\left(\rho_{, \mu}+\gamma_{, \mu}\right)-\left\{\left(1-\mu^{2}\right)\left[1+r \gamma_{, r}\right]^{2}+\left[\mu-\left(1-\mu^{2}\right) \gamma_{, \mu}\right]^{2}\right\}^{-1} \left[ \frac{1}{2}\left[r^2\left[ \gamma_{,rr}+\gamma_{,r}^2\right]\right.\right.\\
&\left.
 -\left(1-\mu^{2}\right)\left( \gamma_{, \mu\mu}+\gamma_{,\mu}^2\right)\right]\left[-\mu+\left(1-\mu^{2}\right) \gamma_{, \mu}\right]+r\gamma_{,r}\left[\frac{1}{2}\mu+\mu r\gamma_r+\frac{1}{2}\left(1-\mu^2\right)\gamma_{,\mu}\right]+\\
 &\left.+\frac{3}{2}\gamma_{,\mu}\left[-\mu^2+\mu(1-\mu^2)\gamma_{,\mu}\right]-r\left(1-\mu^2\right)\left(\gamma_{,r\mu}+\gamma_{,r}\gamma_{,\mu}\right)\left(1+r\gamma_{,r}\right) 
 -\frac{1}{4}\mu r^2\left(\rho_{,r}+\gamma_{,r}\right)^2-\frac{r}{2}\left(1-\mu^2\right)\left(\rho_{,r}+\gamma_{,r}\right)\left(\rho_{,\mu}+\gamma_{,\mu}\right)\right.\\
 &\left.+\frac{1}{4}\mu\left(1-\mu^2\right)\left( \rho_{, \mu}+\gamma_{,\mu}\right)^2-\frac{r^2}{2}\left(1-\mu^2\right)\gamma_{,r}\left(\rho_{,r}+\gamma_{,r}\right)\left(\rho_{,\mu}+\gamma_{,\mu}\right)+\frac{1}{4}\left(1-\mu^2\right)\gamma_{,\mu}\left[r^2\left(\rho_{,r}+\gamma_{,r}\right)^2-\left(1-\mu^2\right)\left(\rho_{,\mu}+\gamma_{,\mu}\right)^2\right]\right.\\
&\left.+ \left(1-\mu^2\right)e^{-2\rho}\left\{\frac{1}{4}r^4\mu\omega_{,r}^2+\frac{1}{2}r^3\left(1-\mu^2\right)\omega_{,r}\omega_{,\mu}-\frac{1}{4}r^2\mu\left(1-\mu^2\right)\omega_{,\mu}^2+\frac{1}{2}r^4\left(1-\mu^2\right)\gamma_{,r}\omega_{,r}\omega_{,\mu}\right.\right.\\
&\left.\left.-\frac{1}{4}r^2\left(1-\mu^2\right)\gamma_{,\mu}\left[r^2\omega_{,r}^2-\left(1-\mu^2\right)\omega_{,\mu}^2\right]\right\}\right].
\end{split}
\label{alpa}
\end{equation}

The pure spinning NS sources are also introduced for completeness:


\begin{equation}
\begin{split}
    &S_{\rho}(r,\mu)= e^{\gamma/2}\left[8\pi e^{2\alpha}\left(\rho_0+\rho_i+ p_{NS}\right)\frac{1+v^2}{1-v^2}+r^2(1-\mu^2)e^{-2\rho}\left(\omega_{,r}^2+\frac{(1-\mu^2)}{r^2}\omega^2_{,\mu}\right)+\right.\\
    &\left.\frac{1}{r}\gamma_{r}-\mu\gamma_{,\mu} +\frac{\rho}{2}\left\{16 \pi e^{2\alpha} p_{NS}-\gamma_{,r}\left(\frac{1}{2}\gamma_{,r}+\frac{1}{r}\right)-\frac{1}{r^2}\gamma_{,\mu}\left(\frac{1-\mu^2}{2}\gamma_{,\mu}-\mu\right) \right\}\right],
\label{SrNS}
\end{split}
\end{equation}

\vspace{1cm}

\begin{equation}
    S_{\gamma}(r,\mu)=
    e^{\gamma/2}\left[16\pi e^{2\alpha} p_{NS}+\frac{\gamma}{2}\left(16\pi e^{2\alpha} p_{NS}-\frac{1}{2}\gamma_{,r}^2-\frac{1-\mu^2}{2r^2}\gamma_{,\mu}^2\right)   \right]
\label{SgNS}
\end{equation}

\vspace{1cm}

\begin{equation}
\begin{split}
    &S_{\omega}(r,\mu)=
    e^{(\gamma-2\rho)/2}\left[-16\pi e^{2\alpha}\frac{(\Omega-\omega)(\rho_0+\rho_i+ p_{NS})}{1-v^2}+\right.\\
    &\left.\omega\left\{ -8\pi e^{2\alpha}\frac{(1+v^2)(\rho_0+\rho_i)+2v^2 p_{NS}}{1-v^2} -\frac{1}{r}\left(2\rho_{,r}+\frac{1}{2}\gamma_{,r}\right)+\frac{\mu}{r^2}\left(2\rho_{,\mu}+\frac{1}{2}\gamma_{,\mu}\right)+ \right.\right.\\ 
    &\left.\left. \frac{1}{4}(4\rho_{,r}^2-\gamma_{,r}^2)+\frac{1-\mu^2}{4r^2}(4\rho_{,\mu}^2-\gamma_{,\mu}^2)-r^2(1-\mu^2)e^{-2\rho}\left[\omega_{,r}^2+\frac{(1-\mu^2)}{r^2}\omega_{,\mu}^2\right]    \right\}\right] .
\label{SwNS}
\end{split}
\end{equation}
\vspace{1cm}

The equation for the function $\alpha$ in the pure NS case is formally identical to \cref{alpa} for the fermion-boson star.

\section{Equations of interest for the initial guess calculations}
\label{appendixB}


\vspace{1cm}

Given $r = r_e\left(\frac{s}{1-s}\right)$, the differentiation rule for the variable change is

\begin{equation}
\begin{split}
    &\frac{dr}{ds}=r_e\frac{d}{ds}\left(\frac{s}{1-s}\right)=r_e\left(\frac{(1-s)-s(-1)}{(1-s)^2}\right)\\&
    \rightarrow dr=\frac{r_e}{(1-s)^2}ds.
\end{split}
\end{equation}

{\bf Static limit: }
\vspace{1cm}

As explained in \cref{section3B}, a change of coordinates from \cref{metricbs} to the metric \cref{metric1} is needed. Apart from the obvious transformation $\omega=0$, we can perform the following identifications between the metric functions in the static and spinning metrics: 

\begin{equation}
    \begin{split}
      &b= e^{(\rho+\gamma)/2}\rightarrow 2\ln b=\rho+\gamma\\
      &e^{\alpha}d\bar{r}=adr\\
      &e^{\alpha}\bar{r}=r,\hspace{0.3cm}e^{(\gamma-\rho)/2}\bar{r}=r\rightarrow\\
      &\rightarrow e^{(\gamma-\rho)/2}=\frac{r}{\bar{r}}\rightarrow \gamma-\rho=2\ln\frac{r}{\bar{r}}
    \end{split}
\end{equation}
The differential equation for the two radii arises as
\begin{equation}
    e^{\alpha}=\frac{r}{\bar{r}}; \frac{dr}{d\bar{r}}=\frac{e^{\alpha}}{a}\rightarrow\frac{dr}{d\bar{r}}=\frac{r}{\bar{r}}\frac{1}{a}.
    \label{betadiff}
\end{equation}
If we introduce the function
\begin{equation}
    \beta^*=\frac{r}{\Bar{r}},
\end{equation}
we can also take into account that $b$ has to be  normalized by its
asymptotic value to deal with the freedom of the initial condition for this potential. By doing so, we recover the asymptotic behavior matching unity at infinity. We denote it as $b^c$, making the difference between the computational and the coordinate variables,
\begin{equation}
    b^*=\frac{b^c}{b^c_{asympt}}.
\end{equation}

We can obtain $\gamma$ and $\rho$ with
\begin{equation}
    \begin{split}
        &\gamma+\rho=2\ln b^*\\
        &\gamma-\rho=2\ln\beta^*\\
        \hline
        &+\rightarrow \gamma=\ln b^*+\ln\beta^*\\
        &-\rightarrow \rho=\ln b^*-\ln\beta^*
    \end{split}
\end{equation}
or in a simpler way,
\begin{equation}
    \begin{split}
        &\alpha=\log\beta^*\\
        &\gamma=\ln b^*+\alpha\\
        &\rho=\ln b^*-\alpha
    \end{split}
\end{equation}

Since $\beta^*$ is still unknown, we rewrite \cref{betadiff} in terms of our new function by using a straightforward change of variable, 
\begin{equation*}
\begin{split}
   & \frac{d\bar{r}}{dr}=\frac{\bar{r}}{r}a\rightarrow \bar{r}=\beta^*/r\rightarrow d\bar{r}=\frac{d\beta^*r+\beta^* dr}{r^2}\\
   &\rightarrow \frac{d\beta^*}{dr}r+\beta^*=\beta^* a\rightarrow \frac{d\beta*}{\beta^*}=\frac{a-1}{r}dr,
    \end{split}
\end{equation*}
and we find the equation for $\beta^*$,
\begin{equation*}
    \beta^*= \exp\left[\int_{r_0}^{r} \frac{a-1}{r}dr\right].
\end{equation*}
We require $\Bar{r}(r)\rightarrow r$ when $r\rightarrow \infty$ to fix the integration constants. 

\vspace{1cm}

\section{Green functions in the KEH}
\label{appendixC}

\vspace{1cm}

We introduce as an example the $\rho$ equation in the KEH seminal paper, in terms of the three-dimensional Green function previous to the expansion. The integral formula for the potential is written in $(r,\mu)$ variables:
\begin{equation}  \rho\sim\int_0^{\infty}dr'\int_{-1}^1d\mu'\int_0^{2\pi}d\psi'r'^2 S_{\rho}(r',\mu')\frac{1}{|\textbf{r}-\textbf{r'}|},
\label{eqrho_lapl}
\end{equation}

and we show the series expansion for the Green function in their radial and angular parts. The angular part is represented by the Legendre polynomials and associated Legendre functions
\begin{equation}
    \frac{1}{|\textbf{r}-\textbf{r'}|}=\sum_{n=0}^{\infty}f_n^2(r,r')\left[P_n(\cos\theta)P_n(\cos\theta')+2\sum_{m=1}^{n}\frac{(n-m)!}{(n+m)!}P_n^m(\cos\theta)P_n^m(\cos\theta')\cos m(\psi-\psi')\right]
\end{equation}
and the radial part 
\begin{equation}
f_n^{2}(r,r')=
\begin{cases}
      (1/r)(r'/r)^n, &for \hspace{1cm} r'/r\leq 1\\
      (1/r')(r/r')^n, &for \hspace{1cm}r'/r >1 .\\
\end{cases}
 \end{equation}

The equality shown in \cref{eqrho_lapl} is proportional to the integral form of the equation $\Delta\left[\rho e^{\gamma/2}\right]=S_{\rho}^T(r,\mu),$ where we have made use of the three-dimensional Laplacian Green function. We then expand the integral equation in polar and radial series and get the metric function's integral equations shown below. Doing the same procedure for the three equations shown in \ref{equationsourcestot}, we get the following expressions for the elliptic potentials $\rho$, $\gamma$ and $\omega$ \cite{komatsu1989rapidly}:

\begin{equation}
\begin{split}
\rho(s,\mu)=&-e^{-\gamma/2}\sum_{n=0}^{\infty}P_{2n}(\mu)\left[\left(\frac{1-s}{s}\right)^{2n+1}\int_0^{s}\frac{ds's'^{2n}}{(1-s')^{2n+2}}
\int_0^1d\mu'P_{2n}(\mu')\widetilde{S}_{\rho}^T(s',\mu') +\left(\frac{s}{1-s}\right)^{2n}\right.\\&\int_0^1\frac{ds'(1-s')^{2n-1}}{s'^{2n+1}}
\left.\int_0^1d\mu'P_{2n}(\mu')\widetilde{S}_{\rho}^T(s',\mu')\right],
\end{split}
    \label{potrho}
\end{equation}

\begin{equation}
    \begin{split}
        \gamma(s,\mu)=&\frac{-2e^{-\gamma/2}}{\pi}\sum_{n=1}^{\infty}\frac{\sin[(2n-1)\theta]}{(2n-1)\sin\theta}\left[\left(\frac{1-s}{s}\right)^{2n}\int_0^s\frac{ds's'^{2n-1}}{(1-s')^{2n+1}}\int_0^1d\mu'\sin[(2n-1)\theta']\widetilde{S}_{\gamma}^T(s',\mu')\right.\\&
        \left.+\left(\frac{s}{1-s}\right)^{2n-2}\int_s^1\frac{ds'(1-s')^{2n-3}}{s'^{2n-1}}\int_0^1d\mu'\sin[(2n-1)\theta']\widetilde{S}_{\gamma}^T(s',\mu')\right],
    \end{split}
    \label{potgama}
\end{equation}

\begin{equation}
    \begin{split}
       & \hat{\omega}(s,\mu)=-e^{(2\rho-\gamma)/2}\sum_{n=1}^{\infty}\frac{P_{2n-1}^{1}(\mu)}{2n(2n-1)\sin\theta}\left[\left(\frac{1-s}{s}\right)^{2n+1}\int_0^s\frac{ds's'^{2n}}{(1-s)^{2n+2}}\int_0^1d\mu'\sin\theta'P_{2n-1}^1(\mu')\widetilde{S}_{\hat{\omega}}^T(s',\mu')\right.\\&
        \left.+\left(\frac{s}{1-s}\right)^{2n-2}\int_0^1 \frac{ds'(1-s')^{2n-3}}{s'^{2n-1}}\int_0^1d\mu'\sin\theta'P_{2n-1}^1(\mu')\widetilde{S}_{\hat{\omega}}^T(s',\mu')\right],
    \end{split}
    \label{potom}
\end{equation}
with $\hat{\omega}=\bar{r}_e\bar{\omega}$, $\widetilde{S}_{\rho}^T(s,\mu)=\bar{r}^2S_{\rho}^T(s,\mu)$, $\widetilde{S}_{\gamma}^T(s,\mu)=\bar{r}^2S_{\gamma}^T(s,\mu)$ and $\widetilde{S}_{\hat{\omega}}^T(s,\mu)=\bar{r}_e\bar{r}^2S_{\omega}^T(s,\mu)$. $P_n(\mu)$ and $P_n^m(\mu)$ are the Legendre and associated polynomials, respectively. 

The following identities are used to reach a better accuracy,
\begin{equation}
    \begin{split}
        &P_n(\mu)d\mu=\frac{1}{n+1}d\left[\mu P_n(\mu)-P_{n-1}(\mu)\right], \hspace{0.2cm}\text{for }\hspace{0.1cm}n>0,\\&
        \sin[(2n-1)\theta]d\mu=d\left(\frac{\sin[2n\theta]}{4n}-\frac{\sin[2(n-1)\theta]}{4(n-1)}\right)\\&
        \text{for }\hspace{0.1cm}n>1,\\&
        \sin\theta d\mu=d\left[\frac{1}{4}\sin(2\theta)-\frac{1}{2}\theta\right], \hspace{0.2cm}\text{for }\hspace{0.1cm}n=1.
    \end{split}
\end{equation}

\end{widetext}
\end{appendix}
\end{document}